\documentclass[a4paper,11pt]{article}
\pdfoutput=1
\usepackage{jheppub_modified}
\usepackage[T1]{fontenc}

\usepackage{eucal}			
\usepackage{mathrsfs}		
\usepackage[sans]{dsfont}	

\usepackage{array}			
\usepackage{longtable}
\usepackage{caption}

\definecolor{rossoCP3}{cmyk}{0,.88,.77,.40}
\definecolor{verdeCP3}{rgb}{0.09765625, 0.57421875, 0.1015625}
\definecolor{bluCP3}{rgb}{0, 0.23, 0.67}

\newcommand{\eg}{e.g.~}
\newcommand{\ie}{i.e.~}

\newcommand{\Eq}[1]{Eq.~\eqref{#1}}
\newcommand{\Sec}[1]{Sec.~\ref{#1}}
\newcommand{\App}[1]{Appendix~\ref{#1}}

\newcommand{\beq}{\begin{equation}}
\newcommand{\eeq}{\end{equation}}
\newcommand{\bol}[1]{\boldsymbol{#1}}
\newcommand{\uno}{\mathds{1}}
\newcommand{\unom}{I}
\newcommand{\pnt}{\rule[-2mm]{0mm}{6mm}}

\newcommand{\Ord}{\CMcal{O}}	
\newcommand{\Op}{{\cal O}}		
\newcommand{\Lag}{\mathscr{L}}	
\newcommand{\mDM}{m}
\newcommand{\el}{\text{el}}
\newcommand{\inel}{\text{inel}}

\newcommand{\gaDM}{\Gamma_\chi}
\newcommand{\gaDMf}{\Gamma_{\chi 5}}
\newcommand{\gaDMfp}{\Gamma_{\chi (5)}}
\newcommand{\gaN}{\Gamma_N}
\newcommand{\gaNf}{\Gamma_{N 5}}
\newcommand{\gaNfp}{\Gamma_{N (5)}}

\hypersetup{colorlinks, bookmarksopen, bookmarksnumbered,
citecolor=verdeCP3, linkcolor=bluCP3, pdfstartview=FitH, urlcolor=rossoCP3
}
\newcommand{\hhref}[1]{\href{http://arxiv.org/abs/#1}{#1}} 

\title{\boldmath A complete Lorentz-to-Galileo dictionary for direct Dark Matter detection}

\author{Eugenio Del Nobile}
\affiliation{School of Physics \& Astronomy, University of Nottingham, \\ University Park, Nottingham, NG7 2RD, UK}
\emailAdd{eugenio.delnobile@nottingham.ac.uk}

\abstract{We determine the most general non-relativistic theory of DM-nucleon scattering complying with the sole requirement of Lorentz invariance, for spin-$0$ and spin-$1/2$ DM. To do so, we first classify a comprehensive list of amplitude terms encompassing the most general Lorentz-covariant $2$-to-$2$ DM-nucleon scattering amplitude. We then match each term to a Galilean-invariant operator at leading-order in the non-relativistic expansion, for both elastic and inelastic (endothermic and exothermic) scattering. Our complete Lorentz-to-Galileo mapping can be used to promptly determine the non-relativistic DM-nucleon interaction and the associated nuclear form factor for any given Lorentz-invariant DM model. It applies to both renormalizable and non-renormalizable theories (such as effective field theories at all orders), at any order of a perturbative expansion. We use our results to prove that, at leading order, Lorentz invariance does not impose restrictions on the set of $16$ Galilean-invariant operators commonly used to parametrize the non-relativistic DM-nucleon interaction. We also predict the lowest effective-operator dimension at which the non-relativistic operators appear in the effective field theory of a singlet DM particle.}

\begin{document} 
\maketitle
\flushbottom

\section{Introduction}
Direct Dark Matter (DM) search experiments aim at detecting the nuclear recoil of detector nuclei upon scattering with a DM particle. If DM particles are gravitationally bound to the Milky Way halo, hence have speeds of order of few hundred km/s at Earth's location, and are heavier than few GeV, the scattering can occur with a whole nucleus rather than with individual nucleons. In these conditions, the scattering can induce nuclear recoils with energy of few keV or above, at the sensitivity threshold of the experiments. Some experiments even manage to have exceptionally low thresholds, becoming sensitive to DM particles with mass in the hundreds of MeV ballpark.

The energy spectrum of the scattering rate measured by the experiments depends on the specific nature of the DM-nucleon interaction. Each type of interaction gives rise to a specific form of the DM-nucleus scattering cross section, which involves the related nuclear form factor. While the natural framework for describing particle interactions is relativistic, computing the DM-nucleus scattering cross section starting from a theory of DM-nucleon interactions requires resorting to a non-relativistic (NR) framework~\cite{Fan:2010gt, Fitzpatrick:2012ix}. Here, the main ingredients used to describe the interaction are not fields but rather the particle momentum and spin three-vector operators. A NR effective field theory was then constructed in Ref.~\cite{Fitzpatrick:2012ix}, where the NR Galilean-invariant operators built out of these ingredients were endowed with a field-like structure. In the same reference, the DM-nucleus cross section was computed for a selection of phenomenologically relevant NR operators. One is then left with the task of establishing the exhaustive set of operators and their combinations that can be of phenomenological interest, and computing the relative cross section.

So far, two distinct approaches have been taken in the literature. One is to start from specific, relativistic DM models and work out the combination of NR operators describing the interaction. The other is to begin already at the NR level, studying the possible operators that can be written down in this framework. Such operators, for DM with spin $0$ and $1/2$, have been completely classified using Galilean symmetry and encoded in a number of building blocks in Ref.~\cite{Dobrescu:2006au}, whose phenomenology was studied \eg in Refs.~\cite{Fitzpatrick:2012ib, Guo:2013ypa, DelNobile:2013sia, Liang:2013dsa, Anand:2013yka, Gresham:2014vja, Catena:2014uqa, Catena:2014epa, Barello:2014uda, Schneck:2015eqa, Catena:2015uua, Scopel:2015baa, Dent:2015zpa, Gluscevic:2015sqa, Aprile:2017aas, Bishara:2017pfq, Bishara:2017nnn, Liu:2017kmx, Catena:2017xqq, Catena:2018ywo, Kang:2018odb}.

In the first approach, where relativistic models are studied one by one, only few of the NR operators are found in mapping to the NR framework. One may therefore wonder whether the other operators can arise at all in more complicated theories, or in corners of parameter space where the dominant contributions analyzed so far are suppressed. Such operators may give rise to interesting phenomenology and it would thus be relevant to know if they can ever arise in a relativistic model, and if so, in what models. Moreover, not all the NR operators may be generated independently. Some may always appear in certain combinations with others, which then raises the question of whether such combinations are simple accidents or have a subtle motivation. A possible reason could be that Lorentz invariance imposes stronger constraints on the scattering amplitude than the Galilean symmetry of the NR framework. Some of these questions remain in the second approach, where the NR operators are studied regardless of their possible origin in a relativistic model. For instance, this approach allows to study the phenomenology of all NR operators but has no say on possible correlations between them, nor on the possibility that some of these operators may never arise in relativistic theories.

In this work we try to answer these questions. To do so, we provide a complete dictionary between the possible terms arising in a general $2$-to-$2$ DM-nucleon scattering amplitude and the NR operators, assuming exclusively Lorentz invariance of the relativistic interaction. In other words, we find a comprehensive list of amplitude terms encompassing the most general Lorentz-covariant DM-nucleon scattering amplitude, and determine for each term the relative NR operator at leading order in the NR expansion. We do so for DM particles with spin $0$ and $1/2$, and treat both the case of elastic and inelastic scattering, where there is a null, positive (endothermic scattering) or negative (exothermic scattering) mass splitting between the outgoing and the incoming DM particles.

We remain agnostic about the possibility of generating the various amplitude terms in specific models. An alternative approach could be to compute the NR limit of an effective field theory of DM-nucleon interactions. To do so, however, one needs to specify the DM gauge quantum numbers (and to restrict to the case of very heavy DM-nucleon interaction mediators). This analysis was carried out \eg in Refs.~\cite{DelNobile:2013sia, Vecchi:2013iza, Bishara:2016hek} for a gauge singlet.

The parametrization of the scattering amplitude is performed in the very same way the matrix element of a current is parametrized in terms of form factors. Textbook examples of this parametrization are, for instance, the formulation of the QED current matrix element in terms of the charge and magnetic dipole moment form factors, to take into account loop effects in the elastic scattering of a charged particle; or the hadronic current matrix elements, to parametrize the effects of hadron compositeness. In both cases this parametrization is constrained by the symmetries of the underlying theory, by the equations of motion, and by the conservation of the QED current. In the same way, we parametrize here the DM-nucleon scattering amplitude imposing solely Lorentz symmetry and the equations of motion (we do not assume any current conservation, to be general). There is no need to specify an underlying Lagrangian, as the proposed complete parametrization encompasses the scattering amplitude of any Lorentz-invariant Lagrangian. For any specific Lagrangian, the DM-nucleon scattering amplitude can be written as a combination of (a subset of the) terms belonging to our collection. The results found analyzing this collection apply then to any Lorentz-invariant theory.

We then match each amplitude term in our collection to a NR operator by performing a NR expansion of the term in the small DM-nucleus relative speed. Each amplitude term is then uniquely matched to the NR operator whose matrix element equals its NR expression. Our NR expansion is thus simply a Taylor-Laurent expansion of the scattering amplitude, which is just a function of the kinematical variables. A different approach could be to perform a NR expansion of a Lagrangian, instead of the scattering amplitude. This approach, called heavy-particle effective theory, allows in a sense to integrate out the DM particle mass, which is large compared to the typical momentum transfer of a DM-nucleus scattering process, without completely integrating out the DM field~\cite{Hill:2011be, Hill:2013hoa, Hill:2014yka, Berlin:2015njh, Bishara:2016hek, Chen:2018uqz}. This expansion method was applied to the effective field theory of a spin-$1/2$ DM particle, singlet under the Standard Model (SM) gauge group, in Ref.~\cite{Bishara:2016hek}, which found the same leading-order NR matching of an analysis where the expansion was instead performed on the scattering amplitude~\cite{DelNobile:2013sia}, as here.

To make a concrete example, let the scattering amplitude of a spin-$1/2$ DM particle $\chi$ scattering elastically off a nucleon $N$ feature the term $c \, q^\mu q_\mu \, K_\alpha \, \bar{u}_\chi \gamma^\alpha u_\chi \, \bar{u}_N \gamma^5 u_N$, where $c$ is a coefficient and $K$, $q$ are combinations of the nucleon and DM momentum four-vectors, defined in \Eq{P K q} below. This scattering amplitude can be \eg obtained at tree level by considering the effective operator $- i c [\square (\bar{\chi} \gamma^\alpha \chi)] (\bar{N} \gamma^5 \overleftrightarrow{\partial_\alpha} N)$, or $\frac{i c}{2 m_N} [\square \partial_\mu (\bar{\chi} \gamma^\alpha \chi)] (\bar{N} \gamma^\mu \gamma^5 \overleftrightarrow{\partial_\alpha} N)$ upon using the equations of motion, with $m_N$ the nucleon mass. The amplitude term can be factored in two parts: the scalar function $c \, q^\mu q_\mu$, which in the above current analogue corresponds to the form factor, and $K_\alpha \, \bar{u}_\chi \gamma^\alpha u_\chi \, \bar{u}_N \gamma^5 u_N$, which corresponds to the parametrized current matrix element. The latter part, \ie that containing the fermion bilinears and all momentum factors that are contracted with them, is what in general in the following we refer to as \emph{Lorentz structure}. The parametrization of the DM-nucleon scattering amplitude consists in identifying a finite set of Lorentz structures that, when multiplied by model-dependent functions of the few available scalars built out of four-momenta, span the DM-nucleon scattering amplitude of all possible Lorentz-invariant theories. The scalar functions can be computed in any given model, but for our model-independent purposes it suffices to regard them as arbitrary functions of the Lorentz scalars built out of four-momenta. We then match each Lorentz structure in this set to the NR operator whose matrix element equals the structure's NR expression. Going back to our example, a NR expansion of the scattering amplitude returns at leading order $8 i c \mDM m_N q^2 \CMcal{I}_\chi \bol{S}_N \cdot \bol{q}$, with $\mDM$ the DM mass, $\bol{q}$ the momentum transfer three-vector, $q^2$ its square, and $\bol{S}_N$ ($\CMcal{I}_\chi$) the nucleon (DM) spin matrix element of the spin $\bol{s}$ (identity) operator (see \Sec{NR limit}). This expression is finally matched to the NR operator $8 i c \mDM m_N q^2 \bol{s}_N \cdot \bol{q} = 8 c \mDM m_N q^2 \Op_{10}$ (see \Eq{NRbuildingblocks}).

The complete Lorentz-to-Galileo mapping provided here can be used to determine the NR DM-nucleon interaction and the associated nuclear form factor, without the need to perform (almost) any computation. One merely needs to express the relativistic scattering amplitude of a chosen model as a linear combination of our comprehensive set of Lorentz structures. Our dictionary then immediately returns the NR theory describing the DM-nucleon interaction. From there, one can straightforwardly apply the formalism of Refs.~\cite{Fitzpatrick:2012ix, Anand:2013yka} to determine the relevant DM-nucleus scattering cross section (at least for those operators for which the nuclear form factor has been computed). The mapping can be used in both renormalizable and non-renormalizable theories (such as effective field theories at all orders), at any order of a perturbative expansion.

The paper is organized as follows. We start in \Sec{NR building blocks} by summarizing the construction of the NR operators introduced in Ref.~\cite{Dobrescu:2006au} (which we call \emph{building blocks} to distinguish them from all other operators, as explained below). We discuss the properties of the different building blocks and we clarify some subtle points about their ``completeness'', the transverse velocity operator and the $\Op_2$ operator. We end the Section with a comprehensive collection of examples where we provide the NR theory of simple renormalizable high-energy models and of selected effective operators. In \Sec{Lorentz-covariant amplitude} we classify the possible terms entering the DM-nucleon scattering amplitude of Lorentz-invariant theories, for both spin-$0$ and spin-$1/2$ DM, and discuss the restrictions that apply to self-conjugated DM. In \Sec{NR limit} we provide the NR operators associated to each term: our comprehensive Lorentz-to-Galileo dictionary can be found in Table~\ref{Spin-0 table} for spin-$0$ DM and in Table~\ref{Spin-1/2 table} for spin-$1/2$ DM. Finally, we conclude in \Sec{Conclusions}.

\section{Non-relativistic building blocks}
\label{NR building blocks}
The possible NR interaction operators for DM-nucleon elastic scattering were originally classified in Ref.~\cite{Dobrescu:2006au}, for spin-$0$ and spin-$1/2$ DM. The analysis carried out in Ref.~\cite{Dobrescu:2006au} is restricted to the center-of-mass frame, but the classification can be easily made frame-independent by exploiting Galilean invariance, as we show in the following (see \eg Ref.~\cite{Fitzpatrick:2012ix}). The construction involves writing down all possible rotationally- and boost-invariant operators built with the operators corresponding to the available classical kinematical ingredients: the initial and final DM momentum, $\bol{p}$ and $\bol{p}'$ respectively, and the initial and final nucleon momentum, $\bol{k}$ and $\bol{k}'$ respectively. Let us also denote with $m_N$ the nucleon mass, and with $\mDM$ and $\mDM + \delta$ the initial and final DM mass, respectively. $\delta = 0$ yields elastic scattering, while $\delta$ positive or negative yields inelastic endothermic or exothermic scattering, respectively. Momentum conservation implies that there are only three independent combinations of momenta, which can be chosen to be
\begin{align}
\bol{P} &\equiv \bol{p} + \bol{p}' \ ,
&
\bol{K} &\equiv \bol{k} + \bol{k}' \ ,
&
i \bol{q} &\equiv i (\bol{p} - \bol{p}') = i (\bol{k}' - \bol{k}) \ .
\end{align}
This choice is convenient as all these operators are hermitian (hermitian conjugation effectively exchanges the initial and final states~\cite{Fitzpatrick:2012ix}), thus any combination thereof is automatically hermitian. Any non-hermiticity, if present, can be parametrized as an imaginary part to the otherwise real operator coefficient. The mass-splitting parameter $\delta$, for instance, effectively breaks hermiticity at the amplitude level by introducing an asymmetry between initial and final states, thus it always appears multiplied by the imaginary unit as $i \delta$.

NR boost invariance then requires operator construction to adopt combinations of momenta that are (proportional to) velocity differences. For elastic scattering, the only two such combinations are $i \bol{q}$ and the ``elastic'' transverse velocity
\beq
\bol{v}^\perp_\el \equiv \frac{\bol{P}}{2 \mDM} - \frac{\bol{K}}{2 m_N} \ .
\eeq
The generalization for generic $\delta$ is
\beq
\bol{v}^\perp_\inel \equiv \bol{v}^\perp_\el - \frac{\delta}{q^2} \bol{q} \ ,
\eeq
satisfying
\begin{align}
\label{v.q}
\bol{v}^\perp_\inel \cdot \bol{q} = 0 \ ,
&&\text{\ie}
&&
\bol{v}^\perp_\el \cdot \bol{q} = \delta \ .
\end{align}
The two definitions of transverse velocity satisfy
\begin{align}
\label{v^2}
{v^\perp_\el}^2 = v_N^2 - \frac{q^2}{4 \mu_N^2} \ ,
&&&
{v^\perp_\inel}^2 = {v^\perp_\el}^2 - \frac{\delta^2}{q^2} \ ,
\end{align}
with $v_N$ the DM-nucleon relative speed and $\mu_N \equiv \mDM m_N / (\mDM + m_N)$ the DM-nucleon reduced mass. For the scattering to be kinematically allowed the DM mass splitting must satisfy $|\delta| \leqslant \frac{1}{2} \mu_T v^2$ (at least for $\delta < 0$), with $\mu_T$ the DM-nucleus reduced mass and $v$ the DM-nucleus relative speed. $v \sim \Ord(10^{-3})$ (in speed-of-light units) is the NR expansion parameter, and we treat
\begin{align}
\frac{q}{\mu_N}, v_N, v^\perp_\el, v^\perp_\inel \sim \Ord(v) \ ,
&&&
\frac{\delta}{\mu_N} \sim \Ord(v^2) \ .
\end{align}
Notice that $\bol{q} = \bol{p} - \bol{p}'$ is not strictly proportional to a velocity difference for $\delta \neq 0$, but the non boost-invariant correction is subleading for $|\delta| \ll \mDM$~\cite{Barello:2014uda}. At the order of the NR expansion where this effect becomes relevant, $\Ord(v^3)$, we also expect other relativistic corrections that spoil Galilean invariance. However, as explained in \Sec{NR limit}, we truncate the expansion at an order where Galilean invariance is intact.

Operators that depend on the DM and/or nucleon spin can be represented by a generic hermitian matrix acting on spin states of each particle. For spin-$1/2$ particles, due to the Pauli matrices $\bol{\sigma}$ forming, together with the identity matrix $\unom_2$, a basis of $2 \times 2$ hermitian matrices, one can parametrize the interaction operator as a linear combination of $\unom_2$ and $\bol{s} \equiv \bol{\sigma} / 2$. Notice, in fact, that any product of two factors of $\bol{s}$ reduces to the aforementioned linear combination through the identity $\sigma^i \sigma^j = \delta_{ij} \unom_2 + i \varepsilon_{ijk} \sigma^k$. The spin operators, $\bol{s}_\chi$ for a spin-$1/2$ DM and $\bol{s}_N$ for the nucleon, are boost invariant. In the following we treat the cases of spin-$0$ and spin-$1/2$ DM in a unified way, by setting $\bol{s}_\chi \equiv \bol{0}$ for spin-$0$ DM.

The NR operators can be classified by combining the above hermitian and boost-invariant ingredients ($i \bol{q}$, $\bol{v}^\perp_\el$, $\bol{s}_\chi$ and $\bol{s}_N$) in all possible rotationally-invariant ways. For generic $\delta$ one may use $\bol{v}^\perp_\inel$ in place of $\bol{v}^\perp_\el$, as done in Ref.~\cite{Barello:2014uda}, however we prefer to adopt $\bol{v}^\perp_\el$ even for inelastic scattering to make direct contact with the formalism of and the nuclear form factors provided in Refs.~\cite{Fitzpatrick:2012ix, Anand:2013yka}, where elastic scattering was assumed (see below for a more in-depth discussion). In contracting the above vectors, one can use both the $\delta_{ij}$ and $\varepsilon_{ijk}$ $SU(2)$-invariant tensors, which means one can take both scalar products as well as vector products of these vectors. Given that products and contractions of two epsilon tensors return sums of products of Kronecker deltas, however, only operators featuring a single vector product are independent. It was found in Ref.~\cite{Dobrescu:2006au} that, with these rules, one can construct $16$ independent Galilean-invariant building blocks, denoted $\Op_i$ below, each of which can be multiplied by an arbitrary function of the scalar operators $q^2$ and ${v^\perp_\el}^2$, as well as of the non-dynamical constants $m_N$, $\mDM$, $\bol{q} \cdot \bol{v}^\perp_\el = \delta$, coupling coefficients and so on. Notice that, as in Ref.~\cite{Fitzpatrick:2012ix}, we call the $16$ operators $\Op_i$'s \emph{building blocks} to distinguish them from all possible operators (this distinction is not necessary in the majority of the phenomenological analyses, where they are often the only operators taken into account). These building blocks are, following the numbering introduced in Refs.~\cite{Fitzpatrick:2012ix, Anand:2013yka},
\beq
\label{NRbuildingblocks}
\begin{aligned}
\Op_1 &\equiv \uno \ ,
&
\\
\Op_3 &\equiv i \bol{s}_N \cdot (\bol{q} \times \bol{v}^\perp_\el) \ ,
&
\Op_4 &\equiv \bol{s}_\chi \cdot \bol{s}_N \ ,
\\
\Op_5 &\equiv i \bol{s}_\chi \cdot (\bol{q} \times \bol{v}^\perp_\el) \ ,
&
\Op_6 &\equiv (\bol{s}_\chi \cdot \bol{q}) (\bol{s}_N \cdot \bol{q}) \ ,
\\
\Op_7 &\equiv \bol{s}_N \cdot \bol{v}^\perp_\el \ ,
&
\Op_8 &\equiv \bol{s}_\chi \cdot \bol{v}^\perp_\el \ ,
\\
\Op_9 &\equiv i \bol{s}_\chi \cdot (\bol{s}_N \times \bol{q}) \ ,
&
\Op_{10} &\equiv i \bol{s}_N \cdot \bol{q} \ ,
\\
\Op_{11} &\equiv i \bol{s}_\chi \cdot \bol{q} \ ,
&
\Op_{12} &\equiv \bol{v}^\perp_\el \cdot (\bol{s}_\chi \times \bol{s}_N) \ ,
\\
\Op_{13} &\equiv i (\bol{s}_\chi \cdot \bol{v}^\perp_\el) (\bol{s}_N \cdot \bol{q}) \ ,
&
\Op_{14} &\equiv i (\bol{s}_\chi \cdot \bol{q}) (\bol{s}_N \cdot \bol{v}^\perp_\el) \ ,
\\
\Op_{15} &\equiv [\bol{s}_\chi \cdot (\bol{q} \times \bol{v}^\perp_\el)] (\bol{s}_N \cdot \bol{q}) \ ,
&
\Op_{16} &\equiv (\bol{s}_\chi \cdot \bol{v}^\perp_\el) (\bol{s}_N \cdot \bol{v}^\perp_\el) \ ,
\\
\Op_{17} &\equiv i [\bol{s}_\chi \cdot (\bol{q} \times \bol{v}^\perp_\el)] (\bol{s}_N \cdot \bol{v}^\perp_\el) \ .
&
\end{aligned}
\eeq
For spin-$0$ DM we only have the subset of $4$ building blocks not featuring $\bol{s}_\chi$, namely
\begin{align*}
& \Op_1, \Op_3, \Op_7, \Op_{10} & \text{spin-$0$ DM}.
\end{align*}
Notice that the two building blocks that can be obtained by exchanging $\bol{s}_\chi \leftrightarrow \bol{s}_N$ in $\Op_{15}$ and $\Op_{17}$ are not independent from the ones above. In fact, by using $\varepsilon_{i j k} \varepsilon_{i a b} = \delta_{j a} \delta_{k b} - \delta_{j b} \delta_{k a}$ to reduce $(\bol{q} \times \bol{v}^\perp_\el) \cdot [(\bol{s}_1 \times \bol{s}_2) \times \bol{x}]$ to a single cross product in two different ways, we get
\begin{align}
(\bol{s}_\chi \cdot \bol{q}) [\bol{s}_N \cdot (\bol{q} \times \bol{v}^\perp_\el)] &= \Op_{15} - q^2 \Op_{12} - i \delta \Op_9 \ ,
\\
i (\bol{s}_\chi \cdot \bol{v}^\perp_\el) [\bol{s}_N \cdot (\bol{q} \times \bol{v}^\perp_\el)] &= \Op_{17} + {v^\perp_\el}^2 \Op_9 - i \delta \Op_{12} \ ,
\end{align}
obtained by setting $\bol{x} = \bol{q}$ and $\bol{x} = \bol{v}^\perp_\el$ respectively.

Despite some of the above building blocks can be written as a product of two other building blocks, namely
\begin{align}
\Op_{13} = \Op_8 \Op_{10} \ ,
&&
\Op_{14} = \Op_7 \Op_{11} \ ,
&&
\Op_{15} = - \Op_5 \Op_{10} \ ,
&&
\Op_{16} = \Op_7 \Op_8 \ ,
&&
\Op_{17} = \Op_5 \Op_7 \ ,
\end{align}
the associated nuclear form factors are not related in any simple way. In this sense, regarding a building block as a product of two other building blocks has no sensible implication: as an example, every one of the $\Op_i$'s can be regarded as the product of itself with $\Op_1$, without the DM-nucleus scattering cross section featuring necessarily the form factor related to $\Op_1$. Furthermore, some products, such as $\Op_3 \Op_5$, may appear at fist sight to have a more complicated structure than those that can be realized with the building blocks~\eqref{NRbuildingblocks}. However, we remark that they can be easily cast in terms of the building blocks~\eqref{NRbuildingblocks}: for instance, expressing the product of two Levi-Civita tensors as a sum of products of Kronecker deltas we get
\beq
\Op_3 \Op_5 = - q^2 {v^\perp_\el}^2 \Op_4 + {v^\perp_\el}^2 \Op_6 + q^2 \Op_{16} + i \delta (\Op_{13} + \Op_{14}) + \delta^2 \Op_4 \ .
\eeq

The building blocks~\eqref{NRbuildingblocks} naturally split in different categories. Considering that the spatial parity $P$ and time reversal $T$ transformations reverse velocities and three-momenta, while spins are reversed by $T$ but kept unchanged by $P$, we can classify the building blocks according to their $P$ and $T$ quantum numbers:
\begin{align*}
& \Op_1, \Op_3, \Op_4, \Op_5, \Op_6, \Op_{16} & \text{$P$-even and $T$-even},
\\
& \Op_7, \Op_8, \Op_9, \Op_{17} & \text{$P$-odd and $T$-even},
\\
& \Op_{13}, \Op_{14} & \text{$P$-even and $T$-odd},
\\
& \Op_{10}, \Op_{11}, \Op_{12}, \Op_{15} & \text{$P$-odd and $T$-odd}.
\end{align*}
When computing the DM-nucleus cross section, interactions that depend or do not depend on the nucleon spin receive quantitatively different enhancement. It is therefore useful to classify the building blocks as to whether they depend on $\bol{s}_N$ (spin-dependent) or not (spin-independent or coherent):
\begin{align*}
& \Op_1, \Op_5, \Op_8, \Op_{11} & \text{spin-independent},
\\
& \Op_3, \Op_4, \Op_6, \Op_7, \Op_9, \Op_{10}, \Op_{12}, \Op_{13}, \Op_{14}, \Op_{15}, \Op_{16}, \Op_{17} & \text{spin-dependent}.
\end{align*}
Finally, the building blocks can be organized hierarchically according to their NR suppression:
\begin{align*}
& \Op_1, \Op_4 & \sim \Ord(v^0),
\\
& \Op_7, \Op_8, \Op_9, \Op_{10}, \Op_{11}, \Op_{12} & \sim \Ord(v^1),
\\
& \Op_3, \Op_5, \Op_6, \Op_{13}, \Op_{14}, \Op_{16} & \sim \Ord(v^2),
\\
& \Op_{15}, \Op_{17} & \sim \Ord(v^3).
\end{align*}
This of course does not mean that building blocks with different levels of suppression cannot appear together in the same operator at leading order. For instance, a fairly common expression when computing the NR limit of a scattering amplitude is
\beq
(\bol{q} \times \bol{s}_\chi) \cdot (\bol{q} \times \bol{s}_N) = q^2 \Op_4 - \Op_6 \ ,
\eeq
where $\Op_4$ and $\Op_6$ both appear at the same order of the non-perturbative expansion (see \eg the example of DM with magnetic dipole moment in \Sec{Examples}). However, in an operator as the one in \Eq{fermion DM vector mediator} below, describing the interaction of a spin-$1/2$ DM particle with a nucleon mediated by a vector field, the $\Ord(v^0)$ building blocks $\Op_1$ and $\Op_4$ (the standard spin-independent and spin-dependent interactions) naturally dominate unless suppressed by very small coefficients.

The most general interaction operator can be written in terms of the NR building blocks~\eqref{NRbuildingblocks} as
\beq\label{fi Oi}
\sum_i f_i(q^2, {v^\perp_\el}^2) \Op_i \ ,
\eeq
with the $f_i$'s arbitrary functions of $q^2$, ${v^\perp_\el}^2$, and of the non-dynamical constants. Notice that the $f_i$'s are part of the operator, as $q^2$ and ${v^\perp_\el}^2$ are themselves quantum operators. The reason for $\Op_2 \equiv {v^\perp_\el}^2$, as first introduced in Ref.~\cite{Fitzpatrick:2012ix}, being missing among the building blocks~\eqref{NRbuildingblocks}, is that we do not treat it as an independent building block but rather we store all the operator's dependence on ${v^\perp_\el}^2$ in the $f_i$'s: in this sense, $\Op_2 = {v^\perp_\el}^2 \Op_1$.

Unfortunately, the different notations used by Refs.~\cite{Dobrescu:2006au} and~\cite{Fitzpatrick:2012ix} seem to have caused some confusion in the literature. Some authors do not include in their study all independent building blocks because some of these were ignored in Refs.~\cite{Fitzpatrick:2012ix, Anand:2013yka}. The analyses carried out in these latter references are admittedly restricted, for instance, to those operators arising at tree level in field theory models with a DM-nucleon mediator with spin $0$ or $1$. We do not find this to be a sufficient reason to only include some building blocks in a comprehensive and truly model-independent analysis. We reiterate that there exists, in fact, an infinite number of possible operators, reflected by the $f_i$'s being in principle arbitrary functions of $q^2$ and ${v^\perp_\el}^2$. For instance, $\Op_4$ and $q^2 \Op_4$ are different operators, though they employ the same building block $\Op_4$; in the same way, $\Op_6 / q^2$ and $q^4 {v^\perp_\el}^{10} \Op_6$ are different operators, though they employ the same building block $\Op_6$. Despite the possible number of operators is infinite, each operator can be uniquely expressed as a linear combination of the $16$ independent building blocks~\eqref{NRbuildingblocks}, as in \Eq{fi Oi}.

Another source of confusion in the literature is about the nature of $\Op_2$, first introduced in Ref.~\cite{Fitzpatrick:2012ix} where however it was excluded from the analysis of NR operators and form factors. In reporting the list of independent NR building blocks, many authors also include $\Op_2$ along with $\Op_1$. As explained above, despite being different operators, $\Op_2$ is not an independent building block in that it is proportional to $\Op_1$. In this sense, $\Op_2 = {v^\perp_\el}^2 \Op_1$ is not dissimilar from $q^2 \Op_1$. There is only a technical reason why one needs to be more careful with factors of $\bol{v}^\perp_\el$ with respect to factors of $\bol{q}$. Due to momentum-conservation laws, the momentum transfer $\bol{q}$ between a DM particle and a bound nucleon is the same as the momentum transfer between the DM particle and the nucleus hosting the nucleon. In other words, the $\bol{q}$ operator only acts on center-of-mass variables, and is therefore insensitive to the internal nuclear structure. For this reason, the operator $f(q^2) \Op_i$ yields the same squared form factor as $\Op_i$, merely multiplied by a factor $f(q^2)^2$ (we are here deliberately confusing the operator $q^2$ with its matrix element between momentum eigenstates). This is not true for the $\bol{v}^\perp_\el$ operator, which acts on both center-of-mass and internal nuclear variables~\cite{Fitzpatrick:2012ix}. Therefore $\Op_2$, despite differing from $\Op_1$ by a mere multiplicative ${v^\perp_\el}^2$ (operatorial) factor, requires a dedicated analysis to determine the related form factor.

The above discussion may possibly explain why $\Op_2$ was explicitly included by Ref.~\cite{Fitzpatrick:2012ix} in the list of potentially interesting operators, while other similar operators such as $q^2 \Op_1$, or $\Op_1 / q^2$ which is dominant for electrically charged DM particles (see \Eq{millicharged} below), were not. $\Op_2$ was however excluded from the analysis of NR operators and form factors of Refs.~\cite{Fitzpatrick:2012ix, Anand:2013yka}, because it is not generated at leading order of the NR expansion by any relativistic interaction~\cite{Fitzpatrick:2012ix}, at least in the tree-level computations performed so far in the literature. In other words, cancellations between Lorentz-invariant operators have to occur for $\Op_2$ to appear in the NR theory. We will confirm here that this is indeed the case, at any order of a perturbative expansion of any (renormalizable or non-renormalizable) Lorentz-invariant theory, for DM with spin $0$ or $1/2$.

To conclude, let us discuss further our choice of defining the NR building blocks with $\bol{v}^\perp_\el$ rather than $\bol{v}^\perp_\inel$, for generic $\delta$. More in general, this is a choice about presenting our results in terms of $\bol{v}^\perp_\el$ rather than $\bol{v}^\perp_\inel$. This was done to make direct contact with the results of Refs.~\cite{Fitzpatrick:2012ix, Anand:2013yka}, where the nuclear form factors corresponding to some of the operators in \Eq{NRbuildingblocks} were provided. If we did otherwise, all our formulas would have to be expressed back in terms of $\bol{v}^\perp_\el$ before the interaction operator could be matched to the correct form factors to be used; or alternatively, one may appropriately modify some of the form factors to match the $\bol{v}^\perp_\inel$ building blocks, as done \eg in Ref.~\cite{Barello:2014uda}. To avoid this extra step, which would be needed to connect our results to those of Refs.~\cite{Fitzpatrick:2012ix, Anand:2013yka}, we decided to present all calculations in terms of $\bol{v}^\perp_\el$. It is straightforward, however, to express our formulas in terms of $\bol{v}^\perp_\inel$. Let us define, for each of the $\Op_i$'s in \Eq{NRbuildingblocks}, the respective building block $\Op^\inel_i$ by substituting $\bol{v}^\perp_\el$ with $\bol{v}^\perp_\inel$. This yields $\Op^\inel_i = \Op_i$, apart from
\begin{subequations}
\begin{align}
\Op^\inel_7 &= \Op_7 + i \frac{\delta}{q^2} \Op_{10} \ ,
&
\Op^\inel_8 &= \Op_8 + i \frac{\delta}{q^2} \Op_{11} \ ,
\\
\Op^\inel_{12} &= \Op_{12} + i \frac{\delta}{q^2} \Op_9 \ ,
&
\Op^\inel_{13} &= \Op_{13} - i \frac{\delta}{q^2} \Op_6 \ ,
\\
\Op^\inel_{14} &= \Op_{14} - i \frac{\delta}{q^2} \Op_6 \ ,
&
\Op^\inel_{16} &= \Op_{16} + i \frac{\delta}{q^2} (\Op_{13} + \Op_{14}) + \frac{\delta^2}{q^4} \Op_6 \ , 
\\
\Op^\inel_{17} &= \Op_{17} - i \frac{\delta}{q^2} \Op_{15} \ .
&
\end{align}
\end{subequations}
One can then use these equations, or more straightforwardly the inverted relations
\begin{subequations}
\label{inelastic Ops}
\begin{align}
\Op_7 &= \Op^\inel_7 - i \frac{\delta}{q^2} \Op^\inel_{10} \ ,
&
\Op_8 &= \Op^\inel_8 - i \frac{\delta}{q^2} \Op^\inel_{11} \ ,
\\
\Op_{12} &= \Op^\inel_{12} - i \frac{\delta}{q^2} \Op^\inel_9 \ ,
&
\Op_{13} &= \Op^\inel_{13} + i \frac{\delta}{q^2} \Op^\inel_6 \ ,
\\
\Op_{14} &= \Op^\inel_{14} + i \frac{\delta}{q^2} \Op^\inel_6 \ ,
&
\Op_{16} &= \Op^\inel_{16} - i \frac{\delta}{q^2} (\Op^\inel_{13} + \Op^\inel_{14}) + \frac{\delta^2}{q^4} \Op^\inel_6 \ , 
\\
\Op_{17} &= \Op^\inel_{17} + i \frac{\delta}{q^2} \Op^\inel_{15} \ ,
&
\end{align}
\end{subequations}
together with \Eq{v^2}, to express all our results in terms of $\bol{v}^\perp_\inel$.

\subsection{Examples}
\label{Examples}
Before continuing, let us make some examples to connect the NR theory discussed above with the high-energy description of some simple renormalizable DM models and DM effective operators. The NR reduction of the scattering amplitude has been performed in the literature for a variety of models (see \eg Refs.~\cite{Fitzpatrick:2012ix, Fitzpatrick:2012ib, DelNobile:2013sia, Barello:2014uda, Bishara:2016hek}). We provide here the leading-order NR theory of spin-$0$ and spin$1/2$ DM particles interacting with nucleons through scalar, vector and tensor (spin-$2$) mediators, together with that of DM particles interacting with photons via a (tiny) electric charge, a magnetic or electric dipole moment, and an anapole moment. For simplicity we will only treat the case of elastic scattering ($\delta = 0$) and non self-conjugated DM.

A scalar DM particle $\phi$ may interact with nucleons through a scalar mediator $S$ with mass $m_S$ via the Lagrangian
\beq
\Lag = \lambda \, \phi^\dagger \phi \, S + \bar{N} (a \unom_4 + i b \gamma^5) N \, S \ ,
\eeq
with $\lambda$ a parameter with mass-dimension $1$ and $a, b$ dimensionless coefficients. The DM-nucleon scattering amplitude reads at tree level
\beq
\CMcal{P}_S \lambda \, \bar{u}_N (a \unom_4 + i b \gamma^5) u_N \ ,
\eeq
with $\CMcal{P}_S = 1 / (q^\mu q_\mu - m_S^2)$, $q^\mu$ being the four-momentum transfer. One can use the formulas and results in \Sec{NR limit} (see otherwise \eg Refs.~\cite{DelNobile:2013sia, Bishara:2016hek}) to show that the amplitude matches to a NR model described by the operator
\beq
- \frac{2 \lambda}{q^2 + m_S^2} (a m_N \Op_1 - b \Op_{10}) \ ,
\eeq
where $q^\mu q_\mu \simeq - q^2$ in the NR limit. In the notation of \Eq{fi Oi} we have at leading order
\begin{align}
f_1(q^2, {v^\perp_\el}^2) = - \frac{2 \lambda a m_N}{q^2 + m_S^2} \ ,
&&&
f_{10}(q^2, {v^\perp_\el}^2) = \frac{2 \lambda b}{q^2 + m_S^2} \ ,
\end{align}
all other $f_i$'s vanishing. Of course, $\Op_{10}$ is negligible with respect to $\Op_1$ unless $a = 0$ or $b / a$ is sufficiently large to compensate for its NR $q / m_N$ suppression. If $S$ is heavy enough, it can be integrated out yielding the effective Lagrangian
\beq
\Lag = \frac{\lambda}{m_S^2} \phi^\dagger \phi \, \bar{N} (a \unom_4 + i b \gamma^5) N + \dots \ .
\eeq
At leading order we recover the above results with all coefficients truncated at the first order of a $q^\mu q_\mu / m_S^2$ expansion, \eg $\CMcal{P}_S \simeq - 1 / m_S^2$ (contact limit). Notice that taking into account higher-order corrections to $\CMcal{P}_S$ in $f_1$ may be subleading to considering $\Op_{10}$, due to their larger $q$ suppression.\footnote{We thank Brando Bellazzini for pointing this out to us.}

A spin-$1/2$ DM particle $\chi$ may interact with nucleons through the scalar $S$ via the Lagrangian
\beq
\Lag = \bar{\chi} (a \unom_4 + i b \gamma^5) \chi \, S + \bar{N} (c \unom_4 + i d \gamma^5) N \, S \ .
\eeq
The DM-nucleon scattering amplitude reads at tree level
\beq
\CMcal{P}_S \, \bar{u}_\chi (a \unom_4 + i b \gamma^5) u_\chi \, \bar{u}_N (c \unom_4 + i d \gamma^5) u_N \ ,
\eeq
which in the NR limit matches
\beq
- \frac{4}{q^2 + m_S^2} (a c \mDM m_N \Op_1 + b c m_N \Op_{11} - a d \mDM \Op_{10} + b d \Op_6) \ .
\eeq
Once again $\Op_1$ dominates unless suppressed by small or vanishing coefficients. $\Op_{10}$ and $\Op_{11}$ are non-relativistically suppressed, and $\Op_6$ is even more suppressed. Integrating $S$ out yields the effective Lagrangian
\beq
\Lag = \frac{1}{m_S^2} \bar{\chi} (a \unom_4 + i b \gamma^5) \chi \, \bar{N} (c \unom_4 + i d \gamma^5) N + \dots \ ,
\eeq
for which the above formulas hold in the contact limit, namely $1 / (q^2 + m_S^2) \simeq 1 / m_S^2$.

A scalar DM $\phi$ may interact with nucleons through a vector mediator $V^\mu$ with mass $m_V$,
\beq
\Lag = [a \partial_\mu (\phi^\dagger \phi) + i b (\phi^\dagger \overleftrightarrow{\partial_\mu} \phi)] V^\mu + \bar{N} (c \gamma_\mu + d \gamma_\mu \gamma^5) N \, V^\mu \ .
\eeq
The DM-nucleon scattering amplitude reads at tree level
\beq
- \CMcal{P}_V \, (- i a q_\mu + b P_\mu) \, \bar{u}_N (c \gamma^\mu + d \gamma^\mu \gamma^5) u_N \ ,
\eeq
with $\CMcal{P}_V = 1 / (q^\mu q_\mu - m_V^2)$, matching to
\beq
\frac{4}{q^2 + m_V^2} (a d m_N \Op_{10} + b c \mDM m_N \Op_1 - 2 b d \mDM m_N \Op_7) \ .
\eeq
Notice that the $a c$ term of the amplitude vanishes due to the equations of motion. As above, $\Op_1$ dominates unless suppressed by small or vanishing coefficients. If $m_V$ is larger than all other masses and energy scales, one can integrate out $V^\mu$ to obtain the effective Lagrangian
\beq
\Lag = - \frac{1}{m_V^2} [a \partial_\mu (\phi^\dagger \phi) + i b (\phi^\dagger \overleftrightarrow{\partial_\mu} \phi)] \, \bar{N} (c \gamma^\mu + d \gamma^\mu \gamma^5) N + \dots \ ,
\eeq
for which again the above results apply in the contact limit. If, instead, $m_V \ll q$, $\CMcal{P}_V \simeq - 1 / q^2$ and the amplitude is greatly enhanced with respect to the case of a heavy mediator.

The interaction of a spin-$1/2$ DM $\chi$ with nucleons through $V^\mu$ can be described by
\beq
\Lag = \bar{\chi} (a \gamma^\mu + b \gamma^\mu \gamma^5) \chi \, V_\mu + \bar{N} (c \gamma^\mu + d \gamma^\mu \gamma^5) N \, V_\mu \ .
\eeq
The DM-nucleon scattering amplitude reads at tree level
\beq
- \CMcal{P}_V \, \bar{u}_\chi (a \gamma^\mu + b \gamma^\mu \gamma^5) u_\chi \, \bar{u}_N (c \gamma_\mu + d \gamma_\mu \gamma^5) u_N \ ,
\eeq
matching to
\beq\label{fermion DM vector mediator}
\frac{4}{q^2 + m_V^2} (a c \mDM m_N \Op_1 + 2 b c \mDM (m_N \Op_8 - \Op_9) - 2 a d m_N (\mDM \Op_7 + \Op_9) - 4 b d \mDM m_N \Op_4) \ .
\eeq
Here $\Op_1$ dominates along with $\Op_4$, unless suppressed by small or vanishing coefficients. $\Op_{10}$ and $\Op_{11}$ are non-relativistically suppressed, and $\Op_6$ is even more suppressed. Integrating out $V^\mu$ yields the effective Lagrangian
\beq
\Lag = - \frac{1}{m_V^2} \bar{\chi} (a \gamma^\mu + b \gamma^\mu \gamma^5) \chi \, \bar{N} (c \gamma^\mu + d \gamma^\mu \gamma^5) N + \dots \ ,
\eeq
for which the above results apply in the contact limit.

A DM particle with a (tiny) electric charge $Q e$ interacts with nucleons through photon exchange via the Lagrangian
\begin{align}
\Lag &= Q e \, i \! \left( \phi^\dagger \overleftrightarrow{\partial_\mu} \phi \right) \! A^\mu
& \text{for spin-$0$ DM},
\\
\Lag &= Q e \, \bar{\chi} \gamma^\mu \chi \, A_\mu
& \text{for spin-$1/2$ DM},
\end{align}
yielding for the DM-nucleon scattering amplitude
\beq
- Q Q_N e^2 \, \CMcal{P}_\gamma \, \bar{u}_\chi \gamma^\mu u_\chi \, \bar{u}_N \gamma_\mu u_N \ ,
\eeq
with $Q_p = 1$ for the proton and $Q_n = 0$ for the neutron, and $\CMcal{P}_\gamma = 1 / q^\mu q_\mu$. In the NR limit this matches to
\beq\label{millicharged}
4 Q Q_N e^2 \frac{\mDM m_N}{q^2} \Op_1 \ ,
\eeq
where we see that the operator $\Op_1 / q^2$ is relevant.

Interactions of spin-$1/2$ DM particles with photons through a magnetic dipole moment $\mu$, an electric dipole moment $d$ or an anapole moment $a$ are described by the effective Lagrangians
\begin{align}
\Lag &= \frac{\mu}{2} \, \bar{\chi} \sigma^{\mu\nu} \chi \, F_{\mu\nu}
& \text{DM magnetic dipole moment},
\\
\Lag &= \frac{d}{2} \, \bar{\chi} i \sigma^{\mu\nu} \gamma^5 \chi \, F_{\mu\nu}
& \text{DM electric dipole moment},
\\
\Lag &= a \, \bar{\chi} \gamma^\mu \gamma^5 \chi \, \partial^\nu F_{\mu\nu}
& \text{DM anapole moment},
\end{align}
respectively. The respective NR operators describing DM-nucleon scattering are, up to an overall sign~\cite{Fitzpatrick:2012ib, DelNobile:2013sia},
\begin{align}
& 2 e \mu \! \left[ Q_N m_N \Op_1 + 4 Q_N \frac{\mDM m_N}{q^2} \Op_5 + 2 g_N \mDM \left( \Op_4 - \frac{\Op_6}{q^2} \right) \right] & \text{magnetic dipole},
\\
& 8 e d Q_N \frac{\mDM m_N}{q^2} \Op_{11} & \text{electric dipole},
\\
& 4 \mDM a e (2 m_N Q_N \Op_8 - g_N \Op_9) & \text{anapole moment},
\end{align}
where $g_p = 5.59$ and $g_n = -3.83$ are the proton and neutron $g$-factors. One sees that also $\Op_5 / q^2$, $\Op_6 / q^2$, and $\Op_{11} / q^2$ appear as NR operators.

The case of a spin-$0$ or spin-$1/2$ DM particle interacting with SM matter through a massive spin-$2$ mediator, $\CMcal{G}^{\mu\nu}$, coupled to the energy-momentum tensors $T_{\text{SM}, \text{DM}}^{\mu\nu}$ of both sectors, was studied \eg in Ref.~\cite{Carrillo-Monteverde:2018phy}. The effective Lagrangian can be written as
\beq
\Lag = - \frac{1}{\Lambda} (a \, \CMcal{G}_{\mu\nu} T_\text{SM}^{\mu\nu} + b \, \CMcal{G}_{\mu\nu} T_\text{DM}^{\mu\nu}) + \dots \ ,
\eeq
with $\Lambda$ a large enough energy scale. The leading-order NR operator describing DM-nucleon scattering was found to be, for both spin-$0$ and spin-$1/2$ DM,
\beq
\frac{a b \mDM^2 m_N^2}{m_\CMcal{G}^2 \Lambda^2} \left( 3 F_T - \frac{1}{3} F_S \right) \Op_1 \ ,
\eeq
with $F_S$ and $F_T$ the gravitational scalar and tensor form factors of the nucleon, respectively.

\section{General Lorentz-covariant DM-nucleon scattering amplitude}
\label{Lorentz-covariant amplitude}
We now proceed to classifying the possible terms featured in the scattering amplitude of a generic Lorentz-invariant DM model. We remain agnostic about the possibility of generating the various terms in specific models, and simply classify all possible terms compatible with Lorentz invariance. The most general DM-nucleon scattering amplitude can be written as
\beq\label{coeffs}
a \gaN + b \gaNf + c_\mu \gaN^\mu + d_\mu \gaNf^\mu + e_{\mu\nu} \gaN^{\mu\nu} \ ,
\eeq
where we defined the ``hermitian'' nucleon bilinears (in the sense that they are the matrix elements of hermitian operators)
\begin{align}
\gaN &\equiv \bar{u}_N(\bol{k}') u_N(\bol{k}) \ ,
&
\gaNf &\equiv \bar{u}_N(\bol{k}') i \gamma^5 u_N(\bol{k}) \ ,
\\
\gaN^\mu &\equiv \bar{u}_N(\bol{k}') \gamma^\mu u_N(\bol{k}) \ ,
&
\gaNf^\mu &\equiv \bar{u}_N(\bol{k}') \gamma^\mu \gamma^5 u_N(\bol{k}) \ ,
\\
\gaN^{\mu\nu} &\equiv \bar{u}_N(\bol{k}') \sigma^{\mu\nu} u_N(\bol{k}) \ ,
&
\gaNf^{\mu\nu} &\equiv \bar{u}_N(\bol{k}') i \sigma^{\mu\nu} \gamma^5 u_N(\bol{k}) \ .
\end{align}
For brevity, we will denote with $\gaNfp$, $\gaNfp^\mu$, $\gaNfp^{\mu\nu}$ both versions of each bilinear, with and without $\gamma^5$. The generality of the above expression for the amplitude is due to the $16$ matrices
\beq\label{Gamma_i}
\Gamma_i = \{ \unom_4, i \gamma^5, \gamma^\mu, \gamma^\mu \gamma^5, \sigma^{\mu\nu} \} \ ,
\eeq
forming a basis of linear hermitian matrices on the four-spinor vector space, where we defined
\begin{align}
\sigma^{\mu\nu} \equiv \frac{i}{2} [\gamma^\mu, \gamma^\nu] \ ,
&&&
\gamma^5 \equiv - \frac{i}{4!} \varepsilon_{\mu \nu \rho \sigma} \gamma^\mu \gamma^\nu \gamma^\rho \gamma^\sigma = i \, \gamma^0 \gamma^1 \gamma^2 \gamma^3 \ .
\end{align}
Any product of Dirac matrices can be reduced to a linear combination of the $\Gamma_i$'s by using standard formulas, see \eg Ref.~\cite{Pal:2007dc}, which means that any nucleon bilinear can be reduced to the form~\eqref{coeffs}. $\gaNf^{\mu\nu}$, which we only introduced here for future reference, is linearly dependent on $\gaN^{\mu\nu}$ due to
\beq\label{sigma gamma5}
\sigma^{\mu\nu} \gamma^5 = \frac{i}{2} \varepsilon^{\mu\nu\rho\sigma} \sigma_{\rho\sigma} \ .
\eeq

For the amplitude~\eqref{coeffs} to transform properly under the Lorentz group, the coefficients $a, b, c_\mu, d_\mu, e_{\mu\nu}$ should transform as Lorentz tensors of rank $0, 1, 2$ as appropriate. These coefficients must be constructed with the ingredients available in the scattering process, which are the initial and final four-momenta of the DM particle, $p$ and $p'$ respectively, and of the nucleon, $k$ and $k'$ respectively. Energy-momentum conservation, which we impose on the amplitude throughout this work, implies that only three out of four momenta are linearly independent. It is convenient to adopt the following ``hermitian'' combinations (see discussion in the previous Section),
\begin{align}
\label{P K q}
P &\equiv p + p' \ ,
&
K &\equiv k + k' \ ,
&
i q &\equiv i (p - p') = i (k' - k) \ ,
\end{align}
where $q$ is the four-momentum transfer.

All scalar, vector and tensor coefficients entering \Eq{coeffs} are in principle arbitrary functions of all the scalars one can build with the above ingredients, namely
\beq\label{scalars}
P^2, K^2, q^\mu q_\mu, P \cdot K, i P \cdot q \ ,
\eeq
where we denoted the squared four-momentum transfer with $q^\mu q_\mu$ to avoid confusion with the squared three-momentum transfer $q^2$. Notice that $K \cdot q = 0$, whereas $P \cdot q$ only vanishes for $\delta = 0$. These functions can be computed in any given model, but cannot be specified in our model-independent approach: parametrizing a scattering amplitude based solely on Lorentz symmetry can only be done up to one or more arbitrary functions of the independent scalars. These functions correspond, for example, to what in the parametrization of the QED and hadronic currents are called form factors, and depend on the underlying model used to compute the amplitude (see \Sec{Examples} for some simple explicit examples). For instance, they trivially depend on the specific coefficients of the DM-nucleon Lagrangian used to compute the scattering amplitude, which in turn depend on the DM-quark and DM-gluon couplings as established by the chiral expansion~\cite{Cirigliano:2012pq, Menendez:2012tm, Klos:2013rwa, Hill:2014yxa, Hoferichter:2015ipa, Bishara:2016hek, Bishara:2017pfq}. In the following we implicitly assume that the coefficients depend on the scalars~\eqref{scalars} through these unspecified functions, and we only focus on the possible arrangements of four-momenta yielding their Lorentz structure.

The Lorentz structure of the coefficients in \Eq{coeffs} can be obtained by taking all possible suitable products and contractions of four-momenta and possibly the completely anti-symmetric Levi-Civita tensor $\varepsilon^{\mu\nu\rho\sigma}$. Since the product of two Levi-Civita tensors can be expressed as a sum of products of metric tensors, we can restrict ourselves to considering the most general tensor structures one can build with just one occurrence of $\varepsilon^{\mu\nu\rho\sigma}$. Some of the tensor coefficients entering $e_{\mu\nu}$ may in principle also be proportional to the metric tensor, but they do not contribute due to the fact that they are contracted with the anti-symmetric tensor $\gaN^{\mu\nu}$. If the DM has spin $1/2$, the coefficients are themselves DM fermion bilinears, and more in general for arbitrary spin the coefficients contain the polarization tensors of the initial and final DM states.

Application of the equations of motion to the amplitude in \Eq{coeffs} does not simplify the problem of determining the most general form of its scalar, vector and tensor coefficients. In fact, if we eliminate $\gaN^\mu$ and $\gaNf^\mu$ using the Gordon and Gordon-like identities
\begin{align}
\label{EOMs}
i \gaN^{\mu\nu} q_\nu &= 2 m_N \gaN^\mu - K^\mu \gaN \ ,
&
\gaNf^{\mu\nu} K_\nu &= 2 m_N \gaNf^\mu + i q^\mu \gaNf \ ,
\end{align}
we can write \Eq{coeffs} as
\beq
a' \gaN + b' \gaNf + e'_{\mu\nu} \gaN^{\mu\nu} + \frac{d_\mu}{2 m_N} \gaNf^{\mu\nu} K_\nu \ ,
\eeq
which means we must still find the most general form of both the scalar ($a'$ and $b'$), vector ($d_\mu$), and tensor ($e'_{\mu\nu}$) coefficients.

Let us introduce some notation before moving on. We will sometimes use uppercase Latin letters ($A^\mu$, $B^\mu$, etc.) to denote the momenta four-vectors in \Eq{P K q}. When contracting momenta with the Levi-Civita tensor, we will substitute the contracted momenta to the contracted tensor indices, \eg $\varepsilon^{\mu A \nu B} = \varepsilon^{\mu \alpha \nu \beta} A_\alpha B_\beta$. Because we only have three independent momenta, $\varepsilon^{\mu A B C}$ either vanishes or is equal to $\pm \Delta^\mu$ with
\beq\label{Delta}
\Delta^\mu \equiv i \varepsilon^{\mu P K q} \ .
\eeq

\subsection{Spin-$0$ DM}
\label{Spin-0 DM}
If the DM has spin-$0$, its polarization tensor is trivial and the coefficients in \Eq{coeffs} can only depend on the momenta. Their Lorentz structure must be given by suitable multiplications and contractions of four-momenta and possibly the $\varepsilon^{\mu\nu\rho\sigma}$ tensor. In the following we treat the case of complex scalar DM, and postpone to \Sec{Real scalar DM} a discussion on the restrictions that apply for real scalar DM.

\subsubsection{Scalar coefficients}
\label{Spin-0 DM scalar coefficients}
The scalar coefficients are functions of the non-zero scalars listed in \Eq{scalars}. Notice that there are only two dynamical variables, the internal energy and the momentum transfer (or alternatively the scattering angle). These can be parametrized in terms of the Mandelstam variables
\begin{align}
s = \left( \frac{P + K}{2} \right)^2 = \frac{1}{4} (P^2 + K^2 + 2 P \cdot K) \ ,
&&&
t = q^\mu q_\mu \ .
\end{align}
Other scalar combinations return the model parameters such as $m_N$, $\mDM$ and $\delta$. For instance, $i P \cdot q = - i \delta (2 \mDM + \delta)$ is a constant.

\subsubsection{Vector coefficients}
Disregarding an arbitrary multiplicative scalar factor, the only possible vector coefficients are
\beq
P^\mu, K^\mu, i q^\mu, \Delta^\mu \ .
\eeq
This list can be effectively reduced by using the following relations, consequence of the equations of motion:
\begin{align}
\gaN^\mu K_\mu &= 2 m_N \gaN \ ,
&
\gaNf^\mu K_\mu &= 0 \ ,
\\
i \gaN^\mu q_\mu &= 0 \ ,
&
i \gaNf^\mu q_\mu &= 2 m_N \gaNf \ .
\end{align}
We have therefore that $\gaNfp^\mu K_\mu$ and $\gaNfp^\mu q_\mu$ either vanish or can be expressed as functions of $\gaNfp$. Given that the problem of determining all possible amplitude terms featuring $\gaNfp$ has been treated in the previous Section on the scalar coefficients, we can effectively restrict our study of the vector coefficients to those included in the collective vector
\beq\label{Lambda}
\Lambda^\mu \equiv P^\mu, \Delta^\mu \ .
\eeq

\subsubsection{Tensor coefficients}
Again disregarding the arbitrary multiplicative scalar, the possible tensor coefficients are
\beq
P^\mu K^\nu, i P^\mu q^\nu, i K^\mu q^\nu, P^\mu \Delta^\nu, K^\mu \Delta^\nu, i q^\mu \Delta^\nu, \varepsilon^{\mu \nu P K}, i \varepsilon^{\mu \nu P q}, i \varepsilon^{\mu \nu K q} \ .
\eeq
Since the tensor coefficients are ultimately contracted with the anti-symmetric tensor $\gaN^{\mu\nu}$, we include neither the metric tensor nor terms of the form $A^\mu A^\nu$ (nor $\Delta^\mu \Delta^\nu$, which can be however expressed in terms of the metric tensor and $A^\mu B^\nu$). For the same reason we do not bother distinguishing $B^\mu A^\nu$ from $A^\mu B^\nu$, and $\Delta^\mu A^\nu$ from $A^\mu \Delta^\nu$.

As above, it is useful to use the equations of motion in the form of \Eq{EOMs} as well as
\begin{align}
\gaN^{\mu\nu} K_\nu &= i q^\mu \gaN \ ,
&
i \gaNf^{\mu\nu} q_\nu &= - K^\mu \gaNf \ ,
\end{align}
together with $ \varepsilon^{\alpha \beta \mu \nu} {\gaN}_{\mu\nu} = - 2 \gaNf^{\alpha\beta}$ by \Eq{sigma gamma5}. We can thus express the amplitude terms involving some of the above tensor coefficients in terms of Lorentz structures already taken into account in our study of the vector and scalar coefficients. For instance, it is clear that any term of the form $A_\mu B_\nu \gaN^{\mu\nu}$ or $\varepsilon_{\mu \nu A B} \gaN^{\mu\nu} = - 2 A_\mu B_\nu \gaNf^{\mu\nu}$ reduces to cases already treated above. We can therefore effectively restrict the above list of tensor coefficients to the sole term
\beq
P^\mu \Delta^\nu \ .
\eeq

\subsubsection{Real scalar DM}
\label{Real scalar DM}
For a self-conjugated field, particle and anti-particle coincide. Any order of the perturbative expansion of the $S$-matrix element can thus be written as a sum of terms, each of which featuring the construction and destruction operators in the two combinations $: a^\dagger(\bol{p}_2) a(\bol{p}_1) :$ and $: a(\bol{p}_2) a^\dagger(\bol{p}_1) :$, $\bol{p}_1$ and $\bol{p}_2$ being integration variables. Only the first term is present for a non self-conjugated field. The first term is multiplied by a function $g(p_1, p_2)$ of four-momenta (including $k$ and $k'$), which also incorporates the nucleon fermion bilinears, whereas the second is multiplied by $g(- p_1, - p_2)$. So upon integration over $\bol{p}_1$ and $\bol{p}_2$ we obtain for the scattering amplitude
\beq\label{Real scalar}
g(p, p') + g(-p', -p) = g(p, p') (1 + \eta^g) \ ,
\eeq
where we denoted with $\eta^g$ the parity of $g$ under $p \leftrightarrow -p'$ exchange, $g(-p', -p) = \eta^g g(p, p')$. For instance, $i q^\mu$ and $K^\mu$ are even under $p \leftrightarrow -p'$, while $P^\mu$ (and thus also $\Delta^\mu$) is odd. Therefore, all scalars in \Eq{scalars} but $P \cdot K$ are even (remember that $i P \cdot q \propto \delta = 0 $ in this case). Also, $i q_\mu \gaNfp^\mu$ has $\eta^g = +1$ whereas $P_\mu \gaNfp^\mu$ and $\Delta_\mu \gaNfp^\mu$ have $\eta^g = -1$. Therefore, the two latter structures are restricted to appear multiplied by $P \cdot K$, or by a scalar function of $P \cdot K$ with the same parity, for a real scalar. On the other hand, terms like $i q_\mu \gaNfp^\mu$ and $P_\mu \Delta_\nu \gaN^{\mu\nu}$ can only appear multiplied by a function of the scalars in \Eq{scalars} with positive parity.

As an example of how to generate these Lorentz structures, the effective interaction operator $(\partial_\mu \phi^2) \bar{N} \gamma^\mu \gamma^5 N$ induces at tree level a scattering amplitude that can be written as \Eq{Real scalar} with $g(p, p') = - i q_\mu \gaNf^\mu$, a structure with $\eta^g = +1$. The effective operator $i (\phi \overleftrightarrow{\partial_\mu} \phi) \bar{N} \gamma^\mu N$ yields instead $g(p, p') = P_\mu \gaN^\mu$, with parity $\eta^g = -1$. As it is, this structure can thus not enter the theory of a real scalar, as one can see already at the Lagrangian level by noticing that $\phi \overleftrightarrow{\partial_\mu} \phi = 0$. On the other hand, a structure as $(P \cdot K) P_\mu \gaN^\mu$ has even parity and is therefore allowed in the theory of a real scalar, where it could arise at tree level from the effective operator $i [(\partial_\mu \phi) \overleftrightarrow{\partial_\nu} \phi - \phi \overleftrightarrow{\partial_\nu} (\partial_\mu \phi)] (\bar{N} \gamma^\mu \overleftrightarrow{\partial^\nu} N)$. Despite these simple examples only feature tree level amplitudes, we remark that \Eq{Real scalar} also holds at loop level.

\subsection{Spin-$1/2$ DM}
\label{Spin-1/2 DM}
For a spin-$1/2$ DM particle $\chi$, apart from depending on the above ingredients (momenta and Levi-Civita tensor), each coefficient in \Eq{coeffs} is a linear combination of the $\gaDM$, $\gaDMf$, $\gaDM^\mu$, $\gaDMf^\mu$, and $\gaDM^{\mu\nu}$ DM bilinears, defined as
\begin{align}
\gaDM &\equiv \bar{u}_{\chi'}(\bol{p}') u_\chi(\bol{p}) \ ,
&
\gaDMf &\equiv \bar{u}_{\chi'}(\bol{p}') i \gamma^5 u_\chi(\bol{p}) \ ,
\\
\gaDM^\mu &\equiv \bar{u}_{\chi'}(\bol{p}') \gamma^\mu u_\chi(\bol{p}) \ ,
&
\gaDMf^\mu &\equiv \bar{u}_{\chi'}(\bol{p}') \gamma^\mu \gamma^5 u_\chi(\bol{p}) \ ,
\\
\gaDM^{\mu\nu} &\equiv \bar{u}_{\chi'}(\bol{p}') \sigma^{\mu\nu} u_\chi(\bol{p}) \ ,
&
\gaDMf^{\mu\nu} &\equiv \bar{u}_{\chi'}(\bol{p}') i \sigma^{\mu\nu} \gamma^5 u_\chi(\bol{p}) \ .
\end{align}
The $u_\chi$ spinor describes the initial DM particle, with mass $m$, while the $u_{\chi'}$ spinor describes the final DM particle, with mass $m + \delta$. $\gaDMf^{\mu\nu}$ is linearly dependent on the others due to \Eq{sigma gamma5}, and we only introduced it here for future reference. As for the nucleon bilinears, we will denote with $\gaDMfp$, $\gaDMfp^\mu$, $\gaDMfp^{\mu\nu}$ both versions of each DM bilinear, with and without $\gamma^5$.

To determine the most general set of the amplitude coefficients in \Eq{coeffs}, we can proceed as follows. We treat here the case of Dirac DM, see \Sec{Majorana DM} below for a discussion of the restrictions that apply for Majorana DM. We first \emph{contract} the linearly-independent DM bilinears $\gaDMfp$, $\gaDMfp^\mu$, $\gaDM^{\mu\nu}$ with a single Levi-Civita tensor in all possible ways. As commented above, products of multiple Levi-Civita tensors do not return independent structures. This exercise produces
\beq\label{structures}
\gaDMfp \ ,
\gaDMfp^\mu \ ,
\gaDMfp^\alpha \varepsilon^{\mu\nu\rho}_\alpha \ ,
\gaDMfp^{\mu\nu} \ ,
\gaDM^{\mu\alpha} \varepsilon^{\nu\rho\sigma}_\alpha \ .
\eeq
We exploited the fact that, by \Eq{sigma gamma5}, ${\gaDM}_{\alpha\beta} \varepsilon^{\alpha \beta \mu \nu} = - 2 \gaDMf^{\mu\nu}$. Notice that, by construction of the above list, no new structure can be obtained by contracting two free indices. We can now suitably contract these structures with momenta four-vectors, and \emph{multiply} (in the sense of a tensor product) the result with tensors formed by momenta (and $\varepsilon^{\mu\nu\rho\sigma}$ whenever not present already), to obtain the most general rank $0$, $1$ and $2$ tensor coefficients. Given that the latter operation of tensor product can only increase the rank, and we are interested in forming tensors of rank at most $2$, the only tensors we can employ in the product are the vector and tensor coefficients discussed above for the case of spin-$0$ DM, \ie $\Lambda^\mu$ (given in \Eq{Lambda}) and $P^\mu \Delta^\nu$.

Regarding contracting the structures in \Eq{structures} with momenta four-vectors, we can again use the equations of motion to find relations among some of these contractions, so to reduce the number of terms that needs being considered. Direct use of the equations of motion returns the following useful relations, analogous to those already considered for the nucleon:
\begin{align}
\gaDM^\mu P_\mu &= (2 \mDM + \delta) \gaDM \ ,
&
\gaDMf^\mu P_\mu &= - i \delta \gaDMf \ ,
\\
i \gaDM^\mu q_\mu &= - i \delta \gaDM \ ,
&
i \gaDMf^\mu q_\mu &= - (2 \mDM + \delta) \gaDMf \ ,
\\
\gaDM^{\mu\nu} P_\nu &= - i q^\mu \gaDM - i \delta \gaDM^\mu \ ,
&
\gaDMf^{\mu\nu} P_\nu &= (2 \mDM + \delta) \gaDMf^\mu - i q^\mu \gaDMf \ ,
\\
i \gaDM^{\mu\nu} q_\nu &= - (2 \mDM + \delta) \gaDM^\mu + P^\mu \gaDM \ ,
&
i \gaDMf^{\mu\nu} q_\nu &= P^\mu \gaDMf - i \delta \gaDMf^\mu \ .
\end{align}
It is thus clear the only expressions that need attention are those where the only momentum four-vector the bilinears $\gaDMfp^\mu$ and $\gaDMfp^{\mu\nu}$ are contracted with is $K^\mu$, given that contractions with $P^\mu$ and/or $i q^\mu$ reduce to expressions involving lower-rank DM bilinears. Other relations exist, that may be of help in reducing the number of structures to be taken into account, see \eg Ref.~\cite{Lorce:2017isp}, but we do not use them here. The point here being not seeking a minimal, complete set of independent structures (assuming such a thing exists), but rather a set of structures that is large enough to encompass the most general scattering amplitude. The list of Lorentz structures obtained following the above prescription (disregarding the arbitrary dependence of any coefficient on the scalars in \Eq{scalars}) is provided in the following.

\subsubsection{Scalar coefficients}
To obtain the scalar coefficients we can only saturate all free indices of the structures in \Eq{structures} with momenta four-vectors:
\beq
\gaDMfp \ ; \
\gaDMfp^\alpha K_\alpha \ ; \
\gaDMfp^\alpha \Delta_\alpha \ ; \
\gaDM^{\alpha\beta} K_\alpha \Delta_\beta \ .
\eeq
Semi-colons separate terms originating from different structures in \Eq{structures}. As for the nucleon tensor bilinears, contraction of $\gaDMfp^{\alpha\beta}$ with any pair of momenta four-vectors can be cast in terms of $\gaDMfp$ and possibly $\gaDMfp^\mu$, which are considered separately. Here and in the following we therefore disregard this type of terms.

\subsubsection{Vector coefficients}
The structures in \Eq{structures} allow to build the following vector coefficients:
\begin{multline}
\gaDMfp \Lambda_\mu \ ; \
\gaDMfp^\mu \ , \ \gaDMfp^\alpha K_\alpha \Lambda_\mu \ ; \
\gaDMfp^\alpha \varepsilon_{\alpha \mu A B} \ , \ \gaDMfp^\alpha \Delta_\alpha P_\mu \ ;
\\
\gaDMfp^{\alpha\mu} K_\alpha \ ; \
\gaDM^{\alpha\mu} \Delta_\alpha \ , \ \gaDM^{\alpha\beta} K_\alpha \varepsilon_{\beta \mu A B} \ , \ \gaDM^{\alpha\beta} K_\alpha \Delta_\beta P_\mu \ .
\end{multline}
$\varepsilon_{\alpha \mu A B}$ here stands for both $\varepsilon_{\alpha \mu P K}$, $i \varepsilon_{\alpha \mu P q}$, and $i \varepsilon_{\alpha \mu K q}$. Contrary to semi-colons, commas separate terms originating from the same structure in \Eq{structures}.

\subsubsection{Tensor coefficients}
The tensor coefficients that can be built are:
\begin{multline}
\gaDMfp P_\mu \Delta_\nu \ ; \
\gaDMfp^\mu \Lambda_\nu \ , \ \gaDMfp^\alpha K_\alpha P_\mu \Delta_\nu \ ; \
\gaDMfp^\alpha \varepsilon_{\alpha P \mu \nu} \ , \ \gaDMfp^\alpha \varepsilon_{\alpha \mu A B} P_\nu \ ;
\\
\gaDMfp^{\mu\nu} \ , \ \gaDM^{\alpha\mu} K_\alpha \Lambda_\nu \ , \ \gaDMf^{\alpha\mu} K_\alpha P_\nu \ ; \
\gaDM^{\alpha\mu} \varepsilon_{\alpha \nu A B} \ , \ \gaDM^{\alpha\beta} K_\alpha \varepsilon_{\beta P \mu \nu} \ , \ \gaDM^{\alpha\beta} K_\alpha \varepsilon_{\beta \mu A B} P_\nu \ , \ \gaDM^{\alpha\mu} \Delta_\alpha P_\nu \ .
\end{multline}

\subsubsection{Majorana DM}
\label{Majorana DM}
For Majorana DM, not only the $u$ spinor but also the $v$ spinor enters the scattering amplitude, since particle and anti-particle coincide. At any order of perturbation theory the scattering amplitude has the form
\beq
\bar{u}_\chi(\bol{p}') \gamma(p, p') u_\chi(\bol{p}) - \bar{v}_\chi(\bol{p}) \gamma(-p', -p) v_\chi(\bol{p}') \ ,
\eeq
with $\gamma$ a matrix-valued function of the external four-momenta (including $k$ and $k'$) in spinor space. $\gamma$ can take the form of a product of Dirac matrices, momenta four-vectors and nucleon fermion bilinears, with Lorentz indices contracted among all of these ingredients, the result being multiplied by a scalar function of momenta. The minus sign in front of the second term originates from normal-ordering the construction and destruction operators of fermion states, $: a(\bol{p}) a^\dagger(\bol{p}') : = - a^\dagger(\bol{p}') a(\bol{p})$, which instead appear automatically normal-ordered for the first term.

As explained at the beginning of this Section one can write, without using the equations of motion, $\gamma(p, p') = \sum_i g_i(p, p') \Gamma_i$, with the $g_i$'s functions of momenta and the $\Gamma_i$'s the matrices of the complete set in \Eq{Gamma_i}. Denoting with $\eta^g_i$ the parity of $g_i$ under $p \leftrightarrow -p'$ exchange, $g_i(-p', -p) = \eta^g_i g_i(p, p')$, the scattering amplitude can be written as
\beq
\sum_i g_i(p, p') [\bar{u}_\chi(\bol{p}') \Gamma_i u_\chi(\bol{p}) - \eta^g_i \, \bar{v}_\chi(\bol{p}) \Gamma_i v_\chi(\bol{p}')] \ .
\eeq
Using now
\beq
\bar{v}_\chi(\bol{p}) \Gamma_i v_\chi(\bol{p}') = - \eta^\text{C}_i \, \bar{u}(\bol{p}') \Gamma_i u(\bol{p}) \ ,
\eeq
with $\eta^\text{C}_i = 1$ for $\Gamma_i = \unom_4, i \gamma^5, \gamma^\mu \gamma^5$ and $\eta^\text{C}_i = - 1$ for $\Gamma_i = \gamma^\mu, \sigma^{\mu\nu}$, we can finally write the scattering amplitude as
\beq
\sum_i g_i(p, p') (1 + \eta^g_i \eta^\text{C}_i) \, \bar{u}_\chi(\bol{p}') \Gamma_i u_\chi(\bol{p}) \ .
\eeq
Amplitude terms with $\eta^g_i \eta^\text{C}_i = -1$ then vanish, and the scattering amplitude contains only terms with $\eta^g_i \eta^\text{C}_i = +1$. This means for instance that terms like $\gaDM^\mu {\gaNfp}_\mu$, $\gaDM^\alpha K_\alpha \gaNfp$, $\gaDMf^\alpha \Delta_\alpha \gaNfp$, and $\gaDM^{\alpha\mu} K_\alpha {\gaNfp}_\mu$, which are allowed for Dirac DM, can only appear in the scattering amplitude for Majorana DM multiplied by $P \cdot K$, or by another scalar function with negative $\eta^g$ parity. On the other hand, terms like $\gaDMf^\mu {\gaNfp}_\mu$, $\gaDM^\alpha \Delta_\alpha \gaNfp$, $\gaDM^\alpha \varepsilon_{\alpha \mu P K} \gaNfp^\mu$, and $i \gaDM^{\alpha\beta} K_\alpha \varepsilon_{\beta \mu K q} P_\nu \gaNfp^{\mu\nu}$, can only be present multiplied by a scalar function with positive $\eta^g$ parity.

As an example, $\gaDM^\mu {\gaN}_\mu$ and $(P \cdot K) \gaDM \gaN$ are the negative-parity tree-level scattering amplitudes induced by the effective operators $\bar{\chi} \gamma^\mu \chi \, \bar{N} \gamma_\mu N$ and $- (\bar{\chi} \overleftrightarrow{\partial^\mu} \chi) (\bar{N} \overleftrightarrow{\partial_\mu} N)$, respectively, which vanish due to $\bar{\chi} \gamma^\mu \chi = 0$ and $\bar{\chi} \overleftrightarrow{\partial_\mu} \chi = 0$ for a Majorana fermion. On the other hand, the positive-parity term $(P \cdot K) \gaDM^\mu {\gaN}_\mu$ is the tree-level amplitude induced by the effective operator $- (\bar{\chi} \gamma^\mu \overleftrightarrow{\partial^\nu} \chi) (\bar{N} \gamma_\mu \overleftrightarrow{\partial_\nu} N)$, which does not vanish.

The list of structures with positive parity is a follows. Scalar coefficients:
\beq
\gaDMfp \ ; \
\gaDMf^\alpha K_\alpha \ ; \
\gaDM^\alpha \Delta_\alpha \ ; \
\gaDM^{\alpha\beta} K_\alpha \Delta_\beta \ .
\eeq
Vector coefficients:
\beq
\gaDMf^\mu \ , \ \gaDM^\alpha K_\alpha \Lambda_\mu \ ; \
\gaDM^\alpha \varepsilon_{\alpha \mu P A} \ , i \gaDMf^\alpha \varepsilon_{\alpha \mu K q} \ , \ \gaDMf^\alpha \Delta_\alpha P_\mu \ ;
\gaDM^{\alpha\mu} \Delta_\alpha \ , \ \gaDM^{\alpha\beta} K_\alpha \varepsilon_{\beta \mu P A} \ .
\eeq
Tensor coefficients:
\begin{multline}
\gaDMfp P_\mu \Delta_\nu \ ; \
\gaDM^\mu \Lambda_\nu \ , \ \gaDMf^\alpha K_\alpha P_\mu \Delta_\nu \ ; \
\gaDM^\alpha \varepsilon_{\alpha P \mu \nu} \ , \ \gaDMf^\alpha \varepsilon_{\alpha \mu P A} P_\nu \ , \ i \gaDM^\alpha \varepsilon_{\alpha \mu K q} P_\nu \ ;
\\
\gaDM^{\alpha\mu} K_\alpha \Lambda_\nu \ , \ \gaDMf^{\alpha\mu} K_\alpha P_\nu \ ; \
\gaDM^{\alpha\mu} \varepsilon_{\alpha \nu P A} \ , \ \gaDM^{\alpha\beta} K_\alpha \varepsilon_{\beta P \mu \nu} \ , \ i \gaDM^{\alpha\beta} K_\alpha \varepsilon_{\beta \mu K q} P_\nu \ .
\end{multline}

\section{Matching to the non-relativistic theory}
\label{NR limit}
In this Section we match each of the scattering amplitude terms classified above to a NR operator. To do so, we perform a Taylor-Laurent expansion in the small expansion parameter $v$ (the DM-nucleus relative speed), which is allowed given that the scattering amplitude is just a function of the kinematical variables. Notice that the expansion is not a simple Taylor series as, for instance, the propagators of massless particles can cause the appearance of negative powers of the momentum transfer (see \eg the case of DM with an electric charge or with a magnetic or electric dipole moment in \Sec{Examples}). Each amplitude term is then uniquely matched to the NR operator whose matrix element equals its NR expression. As remarked in the previous Section, each Lorentz structure can appear in the scattering amplitude multiplied by a function of the scalar factors~\eqref{scalars}. In computing the NR limit of a scalar function times a Lorentz structure, the function is understood to be truncated at the lowest non-zero order.

The NR expansion of four-momenta is carried out at first order in the particle speed, thus expanding the Lorentz factor as $\gamma \simeq 1$. At this order of the NR expansion the Galilean symmetry is intact. The four-vectors of interest here, defined in Eqs.~\eqref{P K q} and~\eqref{Delta}, are expanded as
\begin{align}
P^\mu &\simeq
\begin{pmatrix}
2 \mDM
\\
\bol{P}
\end{pmatrix}
,
&
K^\mu &\simeq
\begin{pmatrix}
2 m_N
\\
\bol{K}
\end{pmatrix}
,
\\
q^\mu &\simeq
\begin{pmatrix}
q^0
\\
\bol{q}
\end{pmatrix}
,
&
\Delta^\mu &\simeq
-
\begin{pmatrix}
i (\bol{P} \times \bol{K}) \cdot \bol{q}
\\
4 i \mDM m_N (\bol{q} \times \bol{v}^\perp_\el)
\end{pmatrix}
,
\end{align}
with
\beq
q^0 \equiv \frac{\bol{K} \cdot \bol{q}}{2 m_N} = \frac{\bol{P} \cdot \bol{q}}{2 \mDM} - \delta \ .
\eeq
We used here $\varepsilon^{0123} = - \varepsilon_{0123} = 1$.

The NR expression of the fermion bilinears can be obtained by using the following first-order approximation of the four-spinor of a generic spin-$1/2$ particle with mass $M$ and momentum $\bol{Q}$, in the chiral representation:
\beq
u(\bol{Q}) \simeq
\frac{1}{\sqrt{4 M}}
\begin{pmatrix}
(2 M - \bol{Q} \cdot \bol{\sigma}) \xi
\\
(2 M + \bol{Q} \cdot \bol{\sigma}) \xi
\end{pmatrix}
,
\eeq
where $\xi$ is a two-spinor, and we adopted the normalization $\bar{u}(\bol{Q}) u(\bol{Q}) = 2 M$. For the final DM particle, the mass $\mDM + \delta$ can be expanded in powers of $\delta \sim \Ord(v^2)$ consistently with the NR expansion, the result being that the mass splitting $\delta$ does not appear in the expression of the spinor at the considered expansion order. Let us now define, for both the nucleon and the spin-$1/2$ DM particle,
\begin{align}
\CMcal{I} \equiv {\xi'}^\dagger \xi \ ,
&&&
\bol{S} \equiv {\xi'}^\dagger \bol{s} \xi \ .
\end{align}
For the nucleon fermion bilinears we then get, at leading order in each entry,
\begin{subequations}
\label{nucleonbilinears}
\begin{align}
\gaN &\simeq 2 m_N \CMcal{I}_N \ ,
\\
\gaNf &\simeq - 2 i \bol{q} \cdot \bol{S}_N \ ,
\\
\gaN^\mu &\simeq
\begin{pmatrix}
2 m_N \CMcal{I}_N
\\
\bol{K} \CMcal{I}_N - 2 i \bol{q} \times \bol{S}_N
\end{pmatrix}
,
\\
\gaNf^\mu &\simeq
\begin{pmatrix}
2 \bol{K} \cdot \bol{S}_N
\\
4 m_N \bol{S}_N
\end{pmatrix}
,
\\
\gaN^{\mu\nu} &\simeq
\begin{pmatrix}
0 & - i \bol{q} \CMcal{I}_N - 2 \bol{K} \times \bol{S}_N
\\
i \bol{q} \CMcal{I}_N + 2 \bol{K} \times \bol{S}_N
&
4 m_N \, \varepsilon_{i j k} S_N^k
\end{pmatrix}
,
\\
\gaNf^{\mu\nu} &\simeq
\begin{pmatrix}
0
&
- 4 m_N \bol{S}_N
\\
4 m_N \bol{S}_N
&
- i \, \varepsilon_{i j k} q^k \CMcal{I}_N - 2 K^i S_N^j + 2 K^j S_N^i
\end{pmatrix}
,
\end{align}
\end{subequations}
while for the DM bilinears we have
\begin{subequations}
\label{fermionbilinears}
\begin{align}
\gaDM &\simeq 2 \mDM \CMcal{I}_\chi \ ,
\\
\gaDMf &\simeq 2 i \bol{q} \cdot \bol{S}_\chi \ ,
\\
\gaDM^\mu &\simeq
\begin{pmatrix}
2 \mDM \CMcal{I}_\chi
\\
\bol{P} \CMcal{I}_\chi + 2 i \bol{q} \times \bol{S}_\chi
\end{pmatrix}
,
\\
\gaDMf^\mu &\simeq
\begin{pmatrix}
2 \bol{P} \cdot \bol{S}_\chi
\\
4 \mDM \bol{S}_\chi
\end{pmatrix}
,
\\
\gaDM^{\mu\nu} &\simeq
\begin{pmatrix}
0 & i \bol{q} \CMcal{I}_\chi - 2 \bol{P} \times \bol{S}_\chi
\\
- i \bol{q} \CMcal{I}_\chi + 2 \bol{P} \times \bol{S}_\chi
&
4 \mDM \, \varepsilon_{i j k} S_\chi^k
\end{pmatrix}
,
\\
\gaDMf^{\mu\nu} &\simeq
\begin{pmatrix}
0
&
- 4 \mDM \bol{S}_\chi
\\
4 \mDM \bol{S}_\chi
&
i \, \varepsilon_{i j k} q^k \CMcal{I}_\chi - 2 P^i S_\chi^j + 2 P^j S_\chi^i
\end{pmatrix}
.
\end{align}
\end{subequations}
Again we notice that $\delta$ does not appear in these expressions at the considered order of the NR expansion.

\subsection{Scalar factors}
The NR expression of the scalar factors in \Eq{scalars} is
\begin{align}
P^2 &\simeq 4 \mDM^2 \ ,
\\
K^2 &\simeq 4 m_N^2 \ ,
\\
q^\mu q_\mu &\simeq - q^2 \ ,
\\
P \cdot K &\simeq 4 \mDM m_N \ ,
\\
i P \cdot q &\simeq - 2 i \mDM \delta \ .
\end{align}
Notice that, oppositely to $q^2$, no factors of ${v^\perp_\el}^2$ appear at leading order. To obtain a ${v^\perp_\el}^2$ factor one has therefore to engineer a cancellation between leading-order terms, \eg
\beq
\label{singling out v^2}
- \left( \frac{P^\mu}{2 \mDM} - \frac{K^\mu}{2 m_N} \right)^2 \simeq {v^\perp_\el}^2 \ .
\eeq
The NR expression of the Mandelstam variables is
\begin{align}
s \simeq (\mDM + m_N) \! \left( \mDM + m_N + \frac{q^2}{4 \mu_N} + \mu_N {v^\perp_\el}^2 + \delta \right) ,
&&&
t \simeq - q^2 \ ,
\end{align}
where we truncated the expansion of $s$ at $\Ord(v^2)$ rather than at the leading $\Ord(v^0)$ to display its dependence on the dynamical variables $q^2$ and $v^\perp_\el$. As explained in \Sec{Spin-0 DM scalar coefficients}, there are only two dynamical variables: the internal energy, which in the NR limit is parametrized most naturally in terms of the DM-nucleon relative velocity and hence ${v^\perp_\el}^2$, and the momentum transfer $q^2$. The scalar factors are functions of these and of the model parameters $m_N$, $\mDM$ and $\delta$.

In the following, as done so far, we neglect the (in principle arbitrary) dependence of the various amplitude terms on the scalar factors, and only focus on their Lorentz structure.

\subsection{Spin-$0$ DM}
\label{Spin-0 DM NR}
In Table~\ref{Spin-0 table} we list the Lorentz structures one can form with the amplitude coefficients given in \Sec{Spin-0 DM}. For each structure we provide the NR operator it matches to in the NR theory and its spatial-parity and time-reversal quantum numbers. In the last column we indicate the $\eta^g$ parity of each structure (see \Sec{Real scalar DM}): for a real scalar DM, structures with $\eta^g = +1$ ($-1$) can only appear multiplied by a scalar function with positive $\eta^g$ parity (negative $\eta^g$ parity, such as $P \cdot K$). Notice that for a self-conjugated DM field one has to set $\delta = 0$.

\begin{table}[t]
\begin{center}
\small
\begin{tabular}{|>{\pnt} c | c | c | c | c |}
\hline
Lorentz structure & NR operator & $P$ & $T$ & $\eta^g$
\\
\hline
$\gaN$ & $2 m_N \Op_1$ & $+$ & $+$ & $+$
\\
$\gaNf$ & $- 2 \Op_{10}$ & $-$ & $-$ & $+$
\\
\hline
$P_\mu \gaN^\mu$ & $4 \mDM m_N \Op_1$ & $+$ & $+$ & $-$
\\
$P_\mu \gaNf^\mu$ & $- 8 \mDM m_N \Op_7$ & $-$ & $+$ & $-$
\\
\hline
$\Delta_\mu \gaN^\mu$ & $8 \mDM m_N (q^2 \Op_7 + i \delta \Op_{10})$ & $-$ & $+$ & $-$
\\
$\Delta_\mu \gaNf^\mu$ & $16 \mDM m_N^2 \Op_3$ & $+$ & $+$ & $-$
\\
\hline
$P_\mu \Delta_\nu \gaN^{\mu\nu}$ & $32 \mDM^2 m_N^2 (- {v^\perp_\el}^2 \Op_{10} + i \delta \Op_7)$ & $-$ & $-$ & $+$
\\
\hline
\end{tabular}
\caption{\label{Spin-0 table} \emph{The Lorentz structures parametrizing the DM-nucleon scattering amplitude for scalar DM, and the NR operators they match to. The third and fourth columns report the spatial-parity and time-reversal quantum numbers of each structure/operator, respectively. The last column indicates the $\eta^g$ parity of each structure, relevant for a real scalar (see \Sec{Real scalar DM}): each structure can only appear in the scattering amplitude multiplied by a scalar function with the same $\eta^g$ parity (notice also that $\delta = 0$ for self-conjugated DM).}}
\end{center}
\end{table}

All NR building blocks available for spin-$0$ DM, namely $\Op_1$, $\Op_3$, $\Op_7$ and $\Op_{10}$, appear independently (meaning that they can be singled out with an appropriate combination of Lorentz structures). They also all appear at least at leading order, \ie not necessarily suppressed by $q^2$ or ${v^\perp_\el}^2$ (operatorial) factors.

$\Op_2$, alias ${v^\perp_\el}^2 \Op_1$, does not appear at leading order. Since ${v^\perp_\el}^2$ is not generated at leading order by the scalar factors~\eqref{scalars} either, we conclude that $\Op_2$ cannot appear at leading order in a theory of spin-$0$ DM without cancellations. This result is valid at any order of a perturbative expansion and in any renormalizable or non-renormalizable theory. The same holds \eg for the operators ${v^\perp_\el}^2 \Op_3$ and ${v^\perp_\el}^2 \Op_7$, while the operator ${v^\perp_\el}^2 \Op_{10}$ is generated by $P_\mu \Delta_\nu \gaN^{\mu\nu}$.

Using Eqs.~\eqref{v^2} and~\eqref{inelastic Ops} we can express the above NR operators in terms of $\bol{v}^\perp_\inel$ rather than $\bol{v}^\perp_\el$, for instance
\begin{align}
\Delta_\mu \gaN^\mu && \to && 8 \mDM m_N q^2 \Op^\inel_7 \ ,
\\
P_\mu \Delta_\nu \gaN^{\mu\nu} && \to && 32 \mDM^2 m_N^2 (- {v^\perp_\inel}^2 \Op^\inel_{10} + i \delta \Op^\inel_7) \ .
\end{align}

For a scalar DM field $\phi$, neutral under the SM gauge group and interacting with the nucleon $N$ through an effective Lagrangian~\cite{DelNobile:2013sia, DEramo:2014nmf, Duch:2014xda, Hisano:2015bma, Bishara:2016hek, Bishara:2017pfq, Brod:2017bsw, Hisano:2017jmz}, it is easy to guess the lowest-dimensional operators that can produce at tree level the Lorentz structures in Table~\ref{Spin-0 table}, assuming all factors of momenta come from derivatives. For instance, the dimension-$5$ effective operator $\phi^\dagger \phi \, \bar{N} (\gamma^5) N$ yields the amplitude term $\gaNfp$, while the dimension-6 operator $i (\phi^\dagger \overleftrightarrow{\partial_\mu} \phi) \bar{N} \gamma^\mu (\gamma^5) N$ yields $P_\mu \gaNfp^\mu$. We also see that to generate the amplitude term $\Delta_\mu \gaNfp^\mu$ we need at least a dimension-$8$ operator such as $\varepsilon^{\mu\nu\rho\sigma} [\partial_\sigma (\phi^\dagger \overleftrightarrow{\partial_\nu} \phi)] [\bar{N} \gamma_\mu (\gamma^5) \overleftrightarrow{\partial_\rho} N]$. Therefore, while the NR building blocks $\Op_1$ and $\Op_{10}$ can arise already at dimension $5$, $\Op_7$ does not arise below dimension $6$ and $\Op_3$ does not arise below dimension $8$ for a complex scalar. For a real scalar, as explained in \Sec{Real scalar DM}, the Lorentz structures can only appear multiplied by a scalar function with the same $\eta^g$ parity. Therefore, $P_\mu \gaNf^\mu$ and $\Delta_\mu \gaNf^\mu$ cannot appear in the scattering amplitude without being multiplied by a $\eta^g$-odd scalar function, that with the least number of momentum factors being $P \cdot K$. One can then argue that the simplest term giving rise to $\Op_7$ is $(P \cdot K) P_\mu \gaNf^\mu$, which can be derived at tree level from the dimension-$8$ effective operator $i [(\partial_\mu \phi) \overleftrightarrow{\partial_\nu} \phi - \phi \overleftrightarrow{\partial_\nu} (\partial_\mu \phi)] (\bar{N} \gamma^\mu \gamma^5 \overleftrightarrow{\partial^\nu} N)$. Similarly, $\Op_3$ can arise from $(P \cdot K) \Delta_\mu \gaNf^\mu$ which is the matrix element of a dimension-$10$ effective operator. Predicting at what order of an effective theory a given NR building block appears cannot be done within the effective field theory formalism, unless one analyzes all possible operators with increasing dimension, which is of course a daunting task. This is a non-trivial way in which our results can be used.

\subsection{Spin-$1/2$ DM}
\label{Spin-1/2 DM NR}
We list the numerous Lorentz structures one can form with the amplitude coefficients given in \Sec{Spin-1/2 DM} in Table~\ref{Spin-1/2 table}, relegated to \App{Mapping for spin-1/2 DM} to avoid cluttering. Again we provide for each Lorentz structure the NR operator it matches to, and indicate its $P$ and $T$ quantum numbers. We also indicate in the last column the $\eta^g \eta^\text{C}$ parity of each Lorentz structure (see \Sec{Majorana DM}): for Majorana DM, structures with $\eta^g \eta^\text{C} = +1$ ($-1$) can only appear in the scattering amplitude multiplied by a scalar function with positive $\eta^g$ parity (negative $\eta^g$ parity, such as $P \cdot K$). Notice again that for a self-conjugated DM field one has to set $\delta = 0$.

As for spin-$0$ DM, all NR building blocks in \Eq{NRbuildingblocks} appear independently at leading order for spin-$1/2$ DM. The NR building blocks $\Op_4$, $\Op_6$, $\Op_9$, $\Op_{10}$, $\Op_{11}$, $\Op_{12}$, $\Op_{13}$, $\Op_{14}$ can also independently appear multiplied by ${v^\perp_\el}^2$, without cancellations of the leading-order contribution. In particular, $\Op_2 = {v^\perp_\el}^2 \Op_1$ does not appear at leading order for spin-$1/2$ DM, as for spin-$0$ DM, at any order of a perturbative expansion and in any renormalizable or non-renormalizable theory. Notice that the Lorentz structures $\gaDM^\alpha \varepsilon_{\alpha \mu P K} \gaN^\mu$, $i \gaDM^{\alpha\beta} K_\alpha \varepsilon_{\beta \mu P q} P_\nu \gaN^{\mu\nu}$ and $i \gaDM^{\alpha\beta} K_\alpha \varepsilon_{\beta \mu K q} P_\nu \gaN^{\mu\nu}$ are only non-vanishing for inelastic scattering.

In the effective field theory of a Dirac DM field $\chi$, neutral under the SM gauge group and interacting with the nucleon $N$~\cite{DelNobile:2013sia, DEramo:2014nmf, Duch:2014xda, Hisano:2015bma, Bishara:2016hek, Bishara:2017pfq, Brod:2017bsw, Hisano:2017jmz}, the NR building blocks $\Op_1$, $\Op_4$, $\Op_5$, $\Op_6$, and $\Op_{11}$, can appear already at dimension $5$ through electric and magnetic dipole interactions with the photon~\cite{DelNobile:2013sia, Bishara:2016hek, Fitzpatrick:2012ib} (see the examples in \Sec{Examples}). Apart from $\Op_5$, they are also induced by the dimension-$6$ four-fermion effective operators together with $\Op_7$, $\Op_8$, $\Op_9$, $\Op_{10}$, $\Op_{11}$, and $\Op_{12}$. However, the effective theory does not allow to predict the order at which the remaining building blocks are generated, unless one analyzes one by one all effective operators of increasing dimension. On the contrary, as already discussed above, the minimum dimension at which a given building block can appear at tree level in the effective theory can be guessed quite easily in our approach, using the following recipe. The building block of interest can be searched for in Table~\ref{Spin-1/2 table} to select the corresponding Lorentz structures (\ie those matching to a NR operator featuring that building block). If all factors of momenta in the amplitude come from derivatives, effective operators can then be easily built whose tree-level scattering amplitude returns the selected Lorentz structures. This exercise reveals that $\Op_3$, $\Op_{13}$, $\Op_{14}$, $\Op_{15}$, $\Op_{16}$, and $\Op_{17}$ can appear at dimension $8$, $7$, $7$, $9$, $8$, and $9$, respectively. Examples of effective operators matching at tree level to NR operators featuring these building blocks are given in Table~\ref{EFT ops}. Special care is needed for self-conjugated DM, as illustrated in \Sec{Spin-0 DM NR} for a real scalar. For Majorana DM one finds that $\Op_1$, $\Op_4$, $\Op_5$, $\Op_6$, $\Op_7$, $\Op_{12}$, and $\Op_{14}$ cannot appear at tree level below dimension $6$, $6$, $8$, $6$, $8$, $8$, and $9$ of the effective theory, respectively. Again we remark that these conclusions cannot be easily deduced within the framework of the effective field theory, while they are quite straightforward in our approach.

\begin{table}[t]
\begin{center}
\begin{tabular}{>{\rule{0mm}{5mm}} c | c c c}
& Lorentz structure & Effective operator & dimension
\\
\hline
$\Op_3$ & $i \gaDM^\alpha \varepsilon_{\alpha \mu K q} \gaNf^\mu$ & $- i \varepsilon^{\alpha\mu\nu\rho} [\partial_\rho (\bar\chi \gamma_\alpha \chi)] [\bar{N} \gamma_\mu \gamma^5 \overleftrightarrow{\partial_\nu} N]$ & $8$
\\
$\Op_{13}$ & $\gaDMf^\alpha K_\alpha \gaNf$ & $i \, \bar\chi \gamma^\alpha \gamma^5 \chi (\bar{N} \gamma^5 \overleftrightarrow{\partial_\alpha} N)$ & $7$
\\
$\Op_{14}$ & $\gaDMf P_\mu \gaNf^\mu$ & $i (\bar\chi \gamma^5 \overleftrightarrow{\partial_\mu} \chi) \bar{N} \gamma^\mu \gamma^5 N$ & $7$
\\
$\Op_{15}$ & $\gaDMf^\alpha \Delta_\alpha \gaNf$ & $\varepsilon^{\alpha\mu\nu\rho} [\partial_\rho (\bar\chi \gamma_\alpha \gamma^5 \overleftrightarrow{\partial_\mu} \chi)] (\bar{N} \gamma^5 \overleftrightarrow{\partial_\nu} N)$ & $9$
\\
$\Op_{16}$ & $\gaDMf^\alpha K_\alpha P_\mu \gaNf^\mu$ & $- (\bar\chi \gamma^\alpha \gamma^5 \overleftrightarrow{\partial_\mu} \chi) (\bar{N} \gamma^\mu \gamma^5 \overleftrightarrow{\partial_\alpha} N)$ & $8$
\\
$\Op_{17}$ & $\gaDM^{\alpha\beta} K_\alpha \varepsilon_{\beta \mu P K} \gaN^\mu$ & $i \varepsilon^{\beta\mu\nu\rho} (\bar\chi \sigma_{\alpha\beta} \overleftrightarrow{\partial_\nu} \chi)] [(\partial_\alpha \bar{N}) \gamma_\mu \overleftrightarrow{\partial_\rho} N - \bar{N} \gamma_\mu \overleftrightarrow{\partial_\rho} (\partial_\alpha N)]$ & $9$
\end{tabular}
\caption{\label{EFT ops} \emph{Examples of effective operators for Dirac DM matching to NR operators containing a given building block. The building blocks in the first column are those that cannot be obtained from a singlet spin-$1/2$ DM-nucleon effective field theory at dimension $6$ or below. For each NR building block, the second column features a Lorentz structure matching to a NR operator containing that building block, see Table~\ref{Spin-1/2 table}. This Lorentz structure is chosen so to contain the least number of momentum factors. Shown in the third column is the effective operator whose matrix element is given by the second column. Its dimension in the effective theory is provided in the last column.}}
\end{center}
\end{table}

\section{Conclusions}
\label{Conclusions}
Non-relativistic (NR) Milky Way halo DM particles interact with whole nuclei within direct DM detection experiments. Computing the DM-nucleus scattering cross section from a relativistic model of DM-nucleon interactions requires determining the associated NR theory, which can be parametrized in terms of the $16$ Galilean-invariant \emph{building blocks}~\eqref{NRbuildingblocks} for DM with spin $0$ or $1/2$. The approaches taken so far in the literature are to compute the NR theory of selected models of DM-nucleon interactions, or otherwise to study the phenomenology of the NR building blocks regardless of their possible origin in high-energy models. The question remained, whether all the building blocks (and more in general all the possible NR operators) can appear independently, or appear at all. In fact, there may in principle exist some degree of dependency among the different building blocks, possibly dictated by subtle constraints imposed by the Lorentz symmetry of the high-energy theory, which the simple models explored so far were unable to reveal.

To answer this question, we classified in this work a comprehensive list of amplitude terms encompassing the most general Lorentz-covariant $2$-to-$2$ DM-nucleon scattering amplitude, and determined for each of them the relative NR operator at leading order in the NR expansion. We did so for DM particles with spin $0$ and $1/2$, and treated both the case of elastic and inelastic (endothermic and exothermic) scattering. This complete Lorentz-to-Galileo mapping can be used to determine the NR DM-nucleon interaction and the associated nuclear form factor, without the need to perform (almost) any computation. Once the relativistic scattering amplitude is expressed as a linear combination of our comprehensive set of \emph{Lorentz structures}, our dictionary immediately returns the associated NR theory. From there, the formalism of Refs.~\cite{Fitzpatrick:2012ix, Anand:2013yka} to determine the relevant DM-nucleus scattering cross section can be straightforwardly applied. Our mapping can be used with both renormalizable and non-renormalizable theories (such as effective field theories at all orders), at any order of a perturbative expansion. The dictionary itself can be found in Table~\ref{Spin-0 table} for spin-$0$ DM and in Table~\ref{Spin-1/2 table} for spin-$1/2$ DM.

Using this complete dictionary we were able to reach the following conclusions. All $16$ ($4$) NR building blocks~\eqref{NRbuildingblocks} are generated independently at leading order of the NR expansion, for spin-$1/2$ (spin-$0$) DM. This could be seen as a confirmation that Lorentz invariance does not impose further constraints than Galilean invariance at the considered expansion order. This also holds for self-conjugated DM, despite the restrictions that apply to the scattering amplitude in this case.

While all NR building blocks can also appear naturally multiplied by a power of the squared three-momentum transfer $q^2$, not all appear multiplied by powers of the squared transverse velocity ${v^\perp_\el}^2$ without cancellation of the leading-order contribution. In particular, $\Op_2 = {v^\perp_\el}^2 \Op_1$ cannot appear at leading order in a theory of spin-$0$ or spin-$1/2$ DM without cancellations. While this result was known for the simple models studied at tree level in the literature so far, our work proves its validity at any order of a perturbative expansion and for any renormalizable or non-renormalizable Lorentz-invariant theory, including effective field theories at all orders.

The NR matching of the effective field theory of a singlet DM field in terms of the building blocks~\eqref{NRbuildingblocks} was only studied in the literature up to dimension $6$, \eg in Refs.~\cite{DelNobile:2013sia, Bishara:2016hek}. Not all the NR building blocks appear at dimension $6$ or below, but predicting at what order of the effective field theory expansion these operators arise, without examining one by one all effective operators of increasing dimension, is impossible in the effective field theory approach. This can instead be done within our framework. One can first select in Table~\ref{Spin-0 table} and Table~\ref{Spin-1/2 table} the Lorentz structures with the lowest mass dimension which map to the NR operator of interest. It is then easy to infer, assuming all factors of momenta come from derivatives, the effective operators whose matrix element equals those Lorentz structures (see Secs.~\ref{Real scalar DM}, \ref{Majorana DM}, \ref{Spin-0 DM NR} and~\ref{Spin-1/2 DM NR} for some examples). Doing so, we can predict that the building block $\Op_3$ does not arise at tree level below dimension $8$ for complex scalar DM, and dimension $10$ for a real scalar. $\Op_7$, which can arise a dimension $6$ for a complex scalar, does not arise at tree level below dimension $8$ for a real scalar. For Dirac DM, $\Op_3$, $\Op_{13}$, $\Op_{14}$, $\Op_{15}$, $\Op_{16}$, and $\Op_{17}$ can appear at tree level in the effective field theory at dimension $8$, $7$, $7$, $9$, $8$, and $9$, respectively. For Majorana DM, $\Op_1$, $\Op_4$, $\Op_5$, $\Op_6$, $\Op_7$, $\Op_{12}$, and $\Op_{14}$ cannot appear at tree level below dimension $6$, $6$, $8$, $6$, $8$, $8$, and $9$, respectively.

\section*{Acknowledgements}
We thank Brando Bellazzini and Luca Vecchi for useful discussions, and Anne Green for guidance. This work was supported by STFC Grant No.~ST/P000703/1.

\appendix

\section{Mapping for spin-$1/2$ DM}
\label{Mapping for spin-1/2 DM}

Table~\ref{Spin-1/2 table} contains the Lorentz structures one can form with the amplitude coefficients given in \Sec{Spin-1/2 DM}, see \Sec{Spin-1/2 DM NR} for further detail. For each Lorentz structure we indicate the NR operator it matches to, together with its $P$ and $T$ quantum numbers. In the last column we indicate the $\eta^g \eta^\text{C}$ parity, which is relevant for Majorana DM (see \Sec{Majorana DM}. In this case one has to set $\delta = 0$).

\footnotesize

\begin{center}
\begin{longtable}{|>{\pnt} c | c | c | c | c |}
\captionsetup{width=\textwidth}
\caption{\label{Spin-1/2 table} \normalsize \emph{Same as Table~\ref{Spin-0 table} but for spin-$1/2$ DM. The last column reports the $\eta^g \eta^\text{C}$ parity of each structure, relevant for a Majorana fermion (see \Sec{Majorana DM}): each structure with $\eta^g \eta^\text{C} = +1$ ($-1$) can only appear in the scattering amplitude multiplied by a scalar function with $\eta^g = +1$ ($-1$) (notice also that $\delta = 0$ for self-conjugated DM).}}
\\
\hline
Lorentz structure & NR operator & $P$ & $T$ & $\eta^g \eta^\text{C}$
\\
\hline
\endfirsthead
\hline
Lorentz structure & NR operator & $P$ & $T$ & $\eta^g \eta^\text{C}$
\\
\hline
\endhead
$\gaDM \gaN$ & $4 \mDM m_N \Op_1$ & $+$ & $+$ & $+$
\\
$\gaDM \gaNf$ & $- 4 \mDM \Op_{10}$ & $-$ & $-$ & $+$
\\
$\gaDMf \gaN$ & $4 m_N \Op_{11}$ & $-$ & $-$ & $+$
\\
$\gaDMf \gaNf$ & $4 \Op_6$ & $+$ & $+$ & $+$
\\
\hline
$\gaDM^\alpha K_\alpha \gaN$ & $8 \mDM m_N^2 \Op_1$ & $+$ & $+$ & $-$
\\
$\gaDM^\alpha K_\alpha \gaNf$ & $- 8 \mDM m_N \Op_{10}$ & $-$ & $-$ & $-$
\\
$\gaDMf^\alpha K_\alpha \gaN$ & $16 \mDM m_N^2 \Op_8$ & $-$ & $+$ & $+$
\\
$\gaDMf^\alpha K_\alpha \gaNf$ & $- 16 \mDM m_N \Op_{13}$ & $+$ & $-$ & $+$
\\
\hline
$\gaDM^\alpha \Delta_\alpha \gaN$ & $- 16 \mDM m_N^2 (q^2 \Op_8 + i \delta \Op_{11})$ & $-$ & $+$ & $+$
\\
$\gaDM^\alpha \Delta_\alpha \gaNf$ & $16 \mDM m_N (q^2 \Op_{13} - i \delta \Op_6)$ & $+$ & $-$ & $+$
\\
$\gaDMf^\alpha \Delta_\alpha \gaN$ & $32 \mDM^2 m_N^2 \Op_5$ & $+$ & $+$ & $-$
\\
$\gaDMf^\alpha \Delta_\alpha \gaNf$ & $32 \mDM^2 m_N \Op_{15}$ & $-$ & $-$ & $-$
\\
\hline
$\gaDM^{\alpha\beta} K_\alpha \Delta_\beta \gaN$ & $64 \mDM^2 m_N^3 ({v^\perp_\el}^2 \Op_{11} - i \delta \Op_8)$ & $-$ & $-$ & $+$
\\
$\gaDM^{\alpha\beta} K_\alpha \Delta_\beta \gaNf$ & $64 \mDM^2 m_N^2 ({v^\perp_\el}^2 \Op_6 + i \delta \Op_{13})$ & $+$ & $+$ & $+$
\\
\hline
$\gaDM P_\mu \gaN^\mu$ & $8 \mDM^2 m_N \Op_1$ & $+$ & $+$ & $-$
\\
$\gaDM P_\mu \gaNf^\mu$ & $- 16 \mDM^2 m_N \Op_7$ & $-$ & $+$ & $-$
\\
$\gaDMf P_\mu \gaN^\mu$ & $8 \mDM m_N \Op_{11}$ & $-$ & $-$ & $-$
\\
$\gaDMf P_\mu \gaNf^\mu$ & $- 16 \mDM m_N \Op_{14}$ & $+$ & $-$ & $-$
\\
\hline
$\gaDM \Delta_\mu \gaN^\mu$ & $16 \mDM^2 m_N (q^2 \Op_7 + i \delta \Op_{10})$ & $-$ & $+$ & $-$
\\
$\gaDM \Delta_\mu \gaNf^\mu$ & $32 \mDM^2 m_N^2 \Op_3$ & $+$ & $+$ & $-$
\\
$\gaDMf \Delta_\mu \gaN^\mu$ & $16 \mDM m_N (q^2 \Op_{14} - i \delta \Op_6)$ & $+$ & $-$ & $-$
\\
$\gaDMf \Delta_\mu \gaNf^\mu$ & $32 \mDM m_N^2 (q^2 \Op_{12} - \Op_{15} + i \delta \Op_9)$ & $-$ & $-$ & $-$
\\
\hline
$\gaDM^\mu {\gaN}_\mu$ & $4 \mDM m_N \Op_1$ & $+$ & $+$ & $-$
\\
$\gaDM^\mu {\gaNf}_\mu$ & $- 8 m_N (\mDM \Op_7 + \Op_9)$ & $-$ & $+$ & $-$
\\
$\gaDMf^\mu {\gaN}_\mu$ & $8 \mDM (m_N \Op_8 - \Op_9)$ & $-$ & $+$ & $+$
\\
$\gaDMf^\mu {\gaNf}_\mu$ & $- 16 \mDM m_N \Op_4$ & $+$ & $+$ & $+$
\\
\hline
$\gaDM^\alpha K_\alpha P_\mu \gaN^\mu$ & $16 \mDM^2 m_N^2 \Op_1$ & $+$ & $+$ & $+$
\\
$\gaDM^\alpha K_\alpha P_\mu \gaNf^\mu$ & $- 32 \mDM^2 m_N^2 \Op_7$ & $-$ & $+$ & $+$
\\
$\gaDMf^\alpha K_\alpha P_\mu \gaN^\mu$ & $32 \mDM^2 m_N^2 \Op_8$ & $-$ & $+$ & $-$
\\
$\gaDMf^\alpha K_\alpha P_\mu \gaNf^\mu$ & $- 64 \mDM^2 m_N^2 \Op_{16}$ & $+$ & $+$ & $-$
\\
\hline
$\gaDM^\alpha K_\alpha \Delta_\mu \gaN^\mu$ & $32 \mDM^2 m_N^2 (q^2 \Op_7 + i \delta \Op_{10})$ & $-$ & $+$ & $+$
\\
$\gaDM^\alpha K_\alpha \Delta_\mu \gaNf^\mu$ & $64 \mDM^2 m_N^3 \Op_3$ & $+$ & $+$ & $+$
\\
$\gaDMf^\alpha K_\alpha \Delta_\mu \gaN^\mu$ & $64 \mDM^2 m_N^2 (q^2 \Op_{16} + i \delta \Op_{13})$ & $+$ & $+$ & $-$
\\
$\gaDMf^\alpha K_\alpha \Delta_\mu \gaNf^\mu$ & $128 \mDM^2 m_N^3 ({v^\perp_\el}^2 \Op_9 + \Op_{17} - i \delta \Op_{12})$ & $-$ & $+$ & $-$
\\
\hline
$\gaDM^\alpha \Delta_\alpha P_\mu \gaN^\mu$ & $- 32 \mDM^2 m_N^2 (q^2 \Op_8 + i \delta \Op_{11})$ & $-$ & $+$ & $-$
\\
$\gaDM^\alpha \Delta_\alpha P_\mu \gaNf^\mu$ & $64 \mDM^2 m_N^2 (q^2 \Op_{16} + i \delta \Op_{14})$ & $+$ & $+$ & $-$
\\
$\gaDMf^\alpha \Delta_\alpha P_\mu \gaN^\mu$ & $64 \mDM^3 m_N^2 \Op_5$ & $+$ & $+$ & $+$
\\
$\gaDMf^\alpha \Delta_\alpha P_\mu \gaNf^\mu$ & $- 128 \mDM^3 m_N^2 \Op_{17}$ & $-$ & $+$ & $+$
\\
\hline
$\gaDM^\alpha \varepsilon_{\alpha \mu P K} \gaN^\mu$ & $- 16 i \mDM m_N \delta \Op_9$ & $-$ & $-$ & $+$
\\
$\gaDM^\alpha \varepsilon_{\alpha \mu P K} \gaNf^\mu$ & $32 \mDM m_N^2 (\Op_{13} - i \delta \Op_4)$ & $+$ & $-$ & $+$
\\
$\gaDMf^\alpha \varepsilon_{\alpha \mu P K} \gaN^\mu$ & $32 \mDM^2 m_N (\Op_{14} - i \delta \Op_4)$ & $+$ & $-$ & $-$
\\
$\gaDMf^\alpha \varepsilon_{\alpha \mu P K} \gaNf^\mu$ & $64 \mDM^2 m_N^2 \Op_{12}$ & $-$ & $-$ & $-$
\\
\hline
$i \gaDM^\alpha \varepsilon_{\alpha \mu P q} \gaN^\mu$ & $8 \mDM [q^2 (m_N \Op_8 - \Op_9) + i m_N \delta \Op_{11}]$ & $-$ & $+$ & $+$
\\
$i \gaDM^\alpha \varepsilon_{\alpha \mu P q} \gaNf^\mu$ & $16 \mDM m_N (\Op_6 - q^2 \Op_4)$ & $+$ & $+$ & $+$
\\
$i \gaDMf^\alpha \varepsilon_{\alpha \mu P q} \gaN^\mu$ & $16 \mDM^2 (- q^2 \Op_4 - m_N \Op_5 + \Op_6)$ & $+$ & $+$ & $-$
\\
$i \gaDMf^\alpha \varepsilon_{\alpha \mu P q} \gaNf^\mu$ & $- 32 \mDM^2 m_N \Op_9$ & $-$ & $+$ & $-$
\\
\hline
\pagebreak
$i \gaDM^\alpha \varepsilon_{\alpha \mu K q} \gaN^\mu$ & $- 8 m_N [q^2 (\mDM \Op_7 + \Op_9) + i \mDM \delta \Op_{10}]$ & $-$ & $+$ & $-$
\\
$i \gaDM^\alpha \varepsilon_{\alpha \mu K q} \gaNf^\mu$ & $16 m_N^2 (- \mDM \Op_3 - q^2 \Op_4 + \Op_6)$ & $+$ & $+$ & $-$
\\
$i \gaDMf^\alpha \varepsilon_{\alpha \mu K q} \gaN^\mu$ & $16 \mDM m_N (- q^2 \Op_4 + \Op_6)$ & $+$ & $+$ & $+$
\\
$i \gaDMf^\alpha \varepsilon_{\alpha \mu K q} \gaNf^\mu$ & $- 32 \mDM m_N^2 \Op_9$ & $-$ & $+$ & $+$
\\
\hline
$\gaDM^{\alpha\mu} K_\alpha {\gaN}_\mu$ & $16 \mDM m_N (\Op_{14} - i \delta \Op_4)$ & $+$ & $-$ & $-$
\\
$\gaDM^{\alpha\mu} K_\alpha {\gaNf}_\mu$ & $8 m_N^2 (- \Op_{10} + 4 \mDM \Op_{12})$ & $-$ & $-$ & $-$
\\
$\gaDMf^{\alpha\mu} K_\alpha {\gaN}_\mu$ & $16 \mDM m_N \Op_9$ & $-$ & $+$ & $-$
\\
$\gaDMf^{\alpha\mu} K_\alpha {\gaNf}_\mu$ & $32 \mDM m_N^2 \Op_4$ & $+$ & $+$ & $-$
\\
\hline
$\gaDM^{\alpha\mu} \Delta_\alpha {\gaN}_\mu$ & $32 \mDM^2 m_N [- m_N {v^\perp_\el}^2 \Op_{11} + q^2 \Op_{12} - \Op_{15} + i \delta (m_N \Op_8 + \Op_9)]$ & $-$ & $-$ & $+$
\\
$\gaDM^{\alpha\mu} \Delta_\alpha {\gaNf}_\mu$ & $64 \mDM^2 m_N^2 (- \Op_{13} + \Op_{14})$ & $+$ & $-$ & $+$
\\
\hline
$\gaDM^{\alpha\beta} K_\alpha \varepsilon_{\beta \mu P K} \gaN^\mu$ & $- 16 \mDM m_N^2 [q^2 \Op_7 + 4 \mDM \Op_{17} + i \delta (\Op_{10} - 4 \mDM \Op_{12})]$ & $-$ & $+$ & $+$
\\
$\gaDM^{\alpha\beta} K_\alpha \varepsilon_{\beta \mu P K} \gaNf^\mu$ & $32 \mDM m_N^3 (- \Op_3 + 4 \mDM {v^\perp_\el}^2 \Op_4 - 4 \mDM \Op_{16})$ & $+$ & $+$ & $+$
\\
\hline
$i \gaDM^{\alpha\beta} K_\alpha \varepsilon_{\beta \mu P q} \gaN^\mu$ & $32 \mDM^2 m_N (- m_N {v^\perp_\el}^2 \Op_{11} + q^2 \Op_{12} - \Op_{15} + i m_N \delta \Op_8)$ & $-$ & $-$ & $+$
\\
$i \gaDM^{\alpha\beta} K_\alpha \varepsilon_{\beta \mu P q} \gaNf^\mu$ & $64 \mDM^2 m_N^2 (\Op_{14} - i \delta \Op_4)$ & $+$ & $-$ & $+$
\\
\hline
$i \gaDM^{\alpha\beta} K_\alpha \varepsilon_{\beta \mu K q} \gaN^\mu$ & $32 \mDM m_N^2 (q^2 \Op_{12} - \Op_{15})$ & $-$ & $-$ & $-$
\\
$i \gaDM^{\alpha\beta} K_\alpha \varepsilon_{\beta \mu K q} \gaNf^\mu$ & $64 \mDM m_N^3 (\Op_{14} - i \delta \Op_4)$ & $+$ & $-$ & $-$
\\
\hline
$\gaDM^{\alpha\beta} K_\alpha \Delta_\beta P_\mu \gaN^\mu$ & $128 \mDM^3 m_N^3 ({v^\perp_\el}^2 \Op_{11} - i \delta \Op_8)$ & $-$ & $-$ & $-$
\\
$\gaDM^{\alpha\beta} K_\alpha \Delta_\beta P_\mu \gaNf^\mu$ & $256 \mDM^3 m_N^3 (- {v^\perp_\el}^2 \Op_{14} + i \delta \Op_{16})$ & $+$ & $-$ & $-$
\\
\hline
$\gaDM P_\mu \Delta_\nu \gaN^{\mu\nu}$ & $64 \mDM^3 m_N^2 (- {v^\perp_\el}^2 \Op_{10} + i \delta \Op_7)$ & $-$ & $-$ & $+$
\\
$\gaDMf P_\mu \Delta_\nu \gaN^{\mu\nu}$ & $64 \mDM^2 m_N^2 ({v^\perp_\el}^2 \Op_6 + i \delta \Op_{14})$ & $+$ & $+$ & $+$
\\
\hline
$\gaDM^\mu P^\nu {\gaN}_{\mu\nu}$ & $16 \mDM m_N (- \Op_{13} + i \delta \Op_4)$ & $+$ & $-$ & $+$
\\
$\gaDM^\mu P^\nu {\gaNf}_{\mu\nu}$ & $- 16 \mDM m_N \Op_9$ & $-$ & $+$ & $+$
\\
$\gaDMf^\mu P^\nu {\gaN}_{\mu\nu}$ & $- 8 \mDM^2 (\Op_{11} + 4 m_N \Op_{12})$ & $-$ & $-$ & $-$
\\
$\gaDMf^\mu P^\nu {\gaNf}_{\mu\nu}$ & $- 32 \mDM^2 m_N \Op_4$ & $+$ & $+$ & $-$
\\
\hline
$\gaDM^\mu \Delta^\nu {\gaN}_{\mu\nu}$ & $32 \mDM m_N^2 (- \mDM {v^\perp_\el}^2 \Op_{10} - \Op_{15} + i \mDM \delta \Op_7)$ & $-$ & $-$ & $+$
\\
$\gaDMf^\mu \Delta^\nu {\gaN}_{\mu\nu}$ & $64 \mDM^2 m_N^2 (- \Op_{13} + \Op_{14})$ & $+$ & $-$ & $-$
\\
\hline
$\gaDM^\alpha K_\alpha P_\mu \Delta_\nu \gaN^{\mu\nu}$ & $128 \mDM^3 m_N^3 (- {v^\perp_\el}^2 \Op_{10} + i \delta \Op_7)$ & $-$ & $-$ & $-$
\\
$\gaDMf^\alpha K_\alpha P_\mu \Delta_\nu \gaN^{\mu\nu}$ & $256 \mDM^3 m_N^3 (- {v^\perp_\el}^2 \Op_{13} + i \delta \Op_{16})$ & $+$ & $-$ & $+$
\\
\hline
$\gaDM^\alpha \varepsilon_{\alpha \mu P K} P_\nu \gaN^{\mu\nu}$ & $- 16 \mDM^2 m_N (q^2 \Op_8 + 4 m_N {v^\perp_\el}^2 \Op_9 + 4 m_N \Op_{17} + i \delta \Op_{11})$ & $-$ & $+$ & $-$
\\
$\gaDMf^\alpha \varepsilon_{\alpha \mu P K} P_\nu \gaN^{\mu\nu}$ & $32 \mDM^3 m_N (- 4 m_N {v^\perp_\el}^2 \Op_4 + \Op_5 + 4 m_N \Op_{16})$ & $+$ & $+$ & $+$
\\
\hline
$i \gaDM^\alpha \varepsilon_{\alpha \mu P q} P_\nu \gaN^{\mu\nu}$ & $32 \mDM^2 m_N (- \Op_{15} + i \delta \Op_9)$ & $-$ & $-$ & $-$
\\
$i \gaDMf^\alpha \varepsilon_{\alpha \mu P q} P_\nu \gaN^{\mu\nu}$ & $64 \mDM^3 m_N (- \Op_{13} + i \delta \Op_4)$ & $+$ & $-$ & $+$
\\
\hline
$i \gaDM^\alpha \varepsilon_{\alpha \mu K q} P_\nu \gaN^{\mu\nu}$ & $32 \mDM m_N^2 [- \mDM {v^\perp_\el}^2 \Op_{10} - \Op_{15} + i \delta (\mDM \Op_7 + \Op_9)]$ & $-$ & $-$ & $+$
\\
$i \gaDMf^\alpha \varepsilon_{\alpha \mu K q} P_\nu \gaN^{\mu\nu}$ & $64 \mDM^2 m_N^2 (- \Op_{13} + i \delta \Op_4)$ & $+$ & $-$ & $-$
\\
\hline
$\gaDM^{\mu\nu} {\gaN}_{\mu\nu}$ & $32 \mDM m_N \Op_4$ & $+$ & $+$ & $-$
\\
$\gaDMf^{\mu\nu} {\gaN}_{\mu\nu}$ & $8 (m_N \Op_{10} - \mDM \Op_{11} - 4 \mDM m_N \Op_{12})$ & $-$ & $-$ & $-$
\\
\hline
$\gaDM^{\alpha\mu} K_\alpha P^\nu {\gaN}_{\mu\nu}$ & $4 \mDM m_N (q^2 \Op_1 + 4 m_N \Op_3 - 16 \mDM m_N {v^\perp_\el}^2 \Op_4 + 4 \mDM \Op_5 + 16 \mDM m_N \Op_{16})$ & $+$ & $+$ & $+$
\\
$\gaDMf^{\alpha\mu} K_\alpha P^\nu {\gaN}_{\mu\nu}$ & $16 \mDM^2 m_N (\Op_{11} + 4 m_N \Op_{12})$ & $-$ & $-$ & $+$
\\
$\gaDM^{\alpha\mu} K_\alpha P^\nu {\gaNf}_{\mu\nu}$ & $16 \mDM m_N^2 (- \Op_{10} + 4 \mDM \Op_{12})$ & $-$ & $-$ & $+$
\\
\hline
$\gaDM^{\alpha\mu} K_\alpha \Delta^\nu {\gaN}_{\mu\nu}$ & $- 32 \mDM m_N^3 (q^2 \Op_7 + 4 \mDM \Op_{17} + i \delta \Op_{10})$ & $-$ & $+$ & $+$
\\
\hline
$\gaDM^{\alpha\mu} \varepsilon^\nu_{\phantom{\nu} \alpha P K} {\gaN}_{\mu\nu}$ & $- 64 \mDM^2 m_N^2 \Op_{12}$ & $-$ & $-$ & $+$
\\
\hline
$i \gaDM^{\alpha\mu} \varepsilon^\nu_{\phantom{\nu} \alpha P q} {\gaN}_{\mu\nu}$ & $32 \mDM^2 m_N \Op_9$ & $-$ & $+$ & $+$
\\
\hline
$i \gaDM^{\alpha\mu} \varepsilon^\nu_{\phantom{\nu} \alpha K q} {\gaN}_{\mu\nu}$ & $32 \mDM m_N^2 \Op_9$ & $-$ & $+$ & $-$
\\
\hline
$\gaDM^{\alpha\beta} K_\alpha \varepsilon_{\beta \mu P K} P_\nu \gaN^{\mu\nu}$ & $64 \mDM^2 m_N^2 [{v^\perp_\el}^2 (- m_N \Op_{10} + \mDM \Op_{11} + 4 \mDM m_N \Op_{12}) + i \delta (m_N \Op_7 - \mDM \Op_8)]$ & $-$ & $-$ & $-$
\\
\hline
$i \gaDM^{\alpha\beta} K_\alpha \varepsilon_{\beta \mu P q} P_\nu \gaN^{\mu\nu}$ & $- 128 i \mDM^3 m_N^2 \delta \Op_{12}$ & $-$ & $+$ & $-$
\\
\hline
$i \gaDM^{\alpha\beta} K_\alpha \varepsilon_{\beta \mu K q} P_\nu \gaN^{\mu\nu}$ & $- 128 i \mDM^2 m_N^3 \delta \Op_{12}$ & $-$ & $+$ & $+$
\\
\hline
$\gaDM^{\alpha\mu} \Delta_\alpha P^\nu {\gaN}_{\mu\nu}$ & $32 \mDM^3 m_N [q^2 \Op_8 + 4 m_N {v^\perp_\el}^2 \Op_9 + 4 m_N \Op_{17} + i \delta (\Op_{11} - 4 m_N \Op_{12})]$ & $-$ & $+$ & $-$
\\
\hline
\end{longtable}
\end{center}

\end{document}